\documentclass[envcountsect,envcountsame,final,orivec]{llncs}
\usepackage[latin1]{inputenc}
\usepackage{hyperref}
\usepackage{amsmath}
\usepackage{amssymb,amsfonts,amsxtra}
\usepackage{stmaryrd}
\usepackage[mathcal,mathscr]{euscript}
\usepackage[english]{babel}
\usepackage{times}

\newdimen\proofrulebreadth \proofrulebreadth=.05em
\newdimen\proofdotseparation \proofdotseparation=1.25ex
\newdimen\proofrulebaseline \proofrulebaseline=2ex
\newcount\proofdotnumber \proofdotnumber=3
\let\then\relax
\def\hfi{\hskip0pt plus.0001fil}
\mathchardef\squigto="3A3B
\newif\ifinsideprooftree\insideprooftreefalse
\newif\ifonleftofproofrule\onleftofproofrulefalse
\newif\ifproofdots\proofdotsfalse
\newif\ifdoubleproof\doubleprooffalse
\let\wereinproofbit\relax
\newdimen\shortenproofleft
\newdimen\shortenproofright
\newdimen\proofbelowshift
\newbox\proofabove
\newbox\proofbelow
\newbox\proofrulename
\def\shiftproofbelow{\let\next\relax\afterassignment\setshiftproofbelow\dimen0 }
\def\shiftproofbelowneg{\def\next{\multiply\dimen0 by-1 }%
\afterassignment\setshiftproofbelow\dimen0 }
\def\setshiftproofbelow{\next\proofbelowshift=\dimen0 }
\def\setproofrulebreadth{\proofrulebreadth}

\def\prooftree{
\ifnum  \lastpenalty=1
\then   \unpenalty
\else   \onleftofproofrulefalse
\fi
\ifonleftofproofrule
\else   \ifinsideprooftree
        \then   \hskip.5em plus1fil
        \fi
\fi
\bgroup
\setbox\proofbelow=\hbox{}\setbox\proofrulename=\hbox{}%
\let\justifies\proofover\let\leadsto\proofoverdots\let\Justifies\proofoverdbl
\let\using\proofusing\let\[\prooftree
\ifinsideprooftree\let\]\endprooftree\fi
\proofdotsfalse\doubleprooffalse
\let\thickness\setproofrulebreadth
\let\shiftright\shiftproofbelow \let\shift\shiftproofbelow
\let\shiftleft\shiftproofbelowneg
\let\ifwasinsideprooftree\ifinsideprooftree
\insideprooftreetrue
\setbox\proofabove=\hbox\bgroup$\displaystyle 
\let\wereinproofbit\prooftree
\shortenproofleft=0pt \shortenproofright=0pt \proofbelowshift=0pt
\onleftofproofruletrue\penalty1
}

\def\eproofbit{
\ifx    \wereinproofbit\prooftree
\then   \ifcase \lastpenalty
        \then   \shortenproofright=0pt  
        \or     \unpenalty\hfil         
        \or     \unpenalty\unskip       
        \else   \shortenproofright=0pt  
        \fi
\fi
\global\dimen0=\shortenproofleft
\global\dimen1=\shortenproofright
\global\dimen2=\proofrulebreadth
\global\dimen3=\proofbelowshift
\global\dimen4=\proofdotseparation
\global\count255=\proofdotnumber
$\egroup  
\shortenproofleft=\dimen0
\shortenproofright=\dimen1
\proofrulebreadth=\dimen2
\proofbelowshift=\dimen3
\proofdotseparation=\dimen4
\proofdotnumber=\count255
}

\def\proofover{
\eproofbit 
\setbox\proofbelow=\hbox\bgroup 
\let\wereinproofbit\proofover
$\displaystyle
}%
\def\proofoverdbl{
\eproofbit 
\doubleprooftrue
\setbox\proofbelow=\hbox\bgroup 
\let\wereinproofbit\proofoverdbl
$\displaystyle
}%
\def\proofoverdots{
\eproofbit 
\proofdotstrue
\setbox\proofbelow=\hbox\bgroup 
\let\wereinproofbit\proofoverdots
$\displaystyle
}%
\def\proofusing{
\eproofbit 
\setbox\proofrulename=\hbox\bgroup 
\let\wereinproofbit\proofusing
\kern0.3em$
}

\def\endprooftree{
\eproofbit 
  \dimen5 =0pt
\dimen0=\wd\proofabove \advance\dimen0-\shortenproofleft
\advance\dimen0-\shortenproofright
\dimen1=.5\dimen0 \advance\dimen1-.5\wd\proofbelow
\dimen4=\dimen1
\advance\dimen1\proofbelowshift \advance\dimen4-\proofbelowshift
\ifdim  \dimen1<0pt
\then   \advance\shortenproofleft\dimen1
        \advance\dimen0-\dimen1
        \dimen1=0pt
        \ifdim  \shortenproofleft<0pt
        \then   \setbox\proofabove=\hbox{%
                        \kern-\shortenproofleft\unhbox\proofabove}%
                \shortenproofleft=0pt
        \fi
\fi
\ifdim  \dimen4<0pt
\then   \advance\shortenproofright\dimen4
        \advance\dimen0-\dimen4
        \dimen4=0pt
\fi
\ifdim  \shortenproofright<\wd\proofrulename
\then   \shortenproofright=\wd\proofrulename
\fi
\dimen2=\shortenproofleft \advance\dimen2 by\dimen1
\dimen3=\shortenproofright\advance\dimen3 by\dimen4
\ifproofdots
\then
        \dimen6=\shortenproofleft \advance\dimen6 .5\dimen0
        \setbox1=\vbox to\proofdotseparation{\vss\hbox{$\cdot$}\vss}%
        \setbox0=\hbox{%
                \advance\dimen6-.5\wd1
                \kern\dimen6
                $\vcenter to\proofdotnumber\proofdotseparation
                        {\leaders\box1\vfill}$%
                \unhbox\proofrulename}%
\else   \dimen6=\fontdimen22\the\textfont2 
        \dimen7=\dimen6
        \advance\dimen6by.5\proofrulebreadth
        \advance\dimen7by-.5\proofrulebreadth
        \setbox0=\hbox{%
                \kern\shortenproofleft
                \ifdoubleproof
                \then   \hbox to\dimen0{%
                        $\mathsurround0pt\mathord=\mkern-6mu%
                        \cleaders\hbox{$\mkern-2mu=\mkern-2mu$}\hfill
                        \mkern-6mu\mathord=$}%
                \else   \vrule height\dimen6 depth-\dimen7 width\dimen0
                \fi
                \unhbox\proofrulename}%
        \ht0=\dimen6 \dp0=-\dimen7
\fi
\let\doll\relax
\ifwasinsideprooftree
\then   \let\VBOX\vbox
\else   \ifmmode\else$\let\doll=$\fi
        \let\VBOX\vcenter
\fi
\VBOX   {\baselineskip\proofrulebaseline \lineskip.2ex
        \expandafter\lineskiplimit\ifproofdots0ex\else-0.6ex\fi
        \hbox   spread\dimen5   {\hfi\unhbox\proofabove\hfi}%
        \hbox{\box0}%
        \hbox   {\kern\dimen2 \box\proofbelow}}\doll%
\global\dimen2=\dimen2
\global\dimen3=\dimen3
\egroup 
\ifonleftofproofrule
\then   \shortenproofleft=\dimen2
\fi
\shortenproofright=\dimen3
\onleftofproofrulefalse
\ifinsideprooftree
\then   \hskip.5em plus 1fil \penalty2
\fi
}

\usepackage{proof}

\usepackage{listings}
\usepackage{amsmath}
\usepackage{threeparttable}

\usepackage{flushend}

\usepackage{url}
\usepackage{pgf}

\begin{document}
\def\Places{{\sf Pl}} 
\def\Loc{{\sf ObjId}} 
\def\Null{\texttt{null}}
\def\gLoc{{\sf grObjId}} 
\def\lLoc{{\sf locObjId}} 
\def\Val{{\sf Val}}
\def\withmath#1{\relax\ifmmode#1\else{$#1$}\fi}
\newcommand{\derives}{\longrightarrow}
\newcommand{\starderives}{\stackrel{\star}{\longrightarrow}}
\def\tuple#1{\withmath{\langle #1 \rangle}}
\def\alt{\withmath{\;\char`\|\;}}
\def\seq{\withmath{\;\char`\;\;}}
\def\eqeq{\;{==}\;}
\def\noteq{\;{!=}\;}
\def\?{\withmath{\;\char`\?\;}}
\def\Hat{{\tt \char`\^}}
\def\OK{\bullet}
\def\VFAIL{v\otimes}
\def\WVFAIL{v\times}
\def\DPE{\mathsf{DP}}
\def\BFE{\mathsf{BF}}
\def\BGE{\mathsf{BG}}
\def\DPFAIL{\DPE\times}
\newcommand\LL[1]{\withmath{\llb#1\rrb}}
\newcommand\csharp{{\sf C\#}}
\newcommand\Implies{\Rightarrow}

\newcommand\rewritesTo[2]{\Longrightarrow_{#1,#2}}

\newcommand\ValS[3]{\texttt{val}\,#1=#2\ #3}
\newcommand\Xten{{\sf X10}}
\newcommand\java{{\sf Java}}
\newcommand\FXten{{\sf FX10}}
\newcommand\new[2]{{\mathit new}\ #1\ #2}
\newcommand\nil{{\mathit nil}}
\newcommand\dom[1]{{\it dom}(#1)}
\newcommand\place[2]{{\mathit place}\ #1\ #2}

\newcommand\var{{\mathit var}}
\newcommand\true{\withmath{\mathit true}}
\newcommand\false{\withmath{\mathit false}}
\newcommand{\defeq}{\stackrel{def}{=}}

\def\llb{\withmath{\lbrack\!\lbrack}}
\def\rrb{\withmath{\rbrack\!\rbrack}}

\newcommand\code[1]{\texttt{#1}}
\newcommand\finsub{{\subseteq_{\mathit fin}}}
\newcommand\atomic[1]{\texttt{atomic}\, #1}
\newcommand\rep[1]{!\; #1}
\newcommand\New[3]{\texttt{new}(#1,#2,#3)}
\newcommand\stuck[1]{\texttt{stuck}\ #1}
\newcommand\Par{{\mathit par}}

\newcommand\Skip{\texttt{skip}}

\newcommand\NULL{\texttt{null}}
\newcommand\GloRef[3]{\mbox{\texttt{globalref}}(#1,#2,#3)}
\newcommand\Copy[3]{{\sf copy}(#1,#2,#3)}

\newcommand\at[2]{\texttt{at}(#1)\ #2}
\newcommand\At[2]{\withmath{\overline{\texttt{at}}}\,(#1)\,#2}
\newcommand\Atttwo[2]{\withmath{\texttt{at}}\,(#1)\,#2}
\newcommand\Att[4]{\texttt{at}(#1)(\texttt{val}\,#2=#3)\,#4}

\newcommand\Var[2]{\texttt{var}\,(#1)=#2}
\newcommand\TryCatch[2]{\texttt{try}\,#1\,\texttt{catch}\,#2}
\newcommand\async[1]{\texttt{async}\ #1}
\newcommand\asynca[1]{\withmath{\overline{\texttt{async}}\,} #1}
\newcommand\isReallyAsync[1]{\texttt{isReallyAsync}\ #1}
\newcommand\isAsync[1]{\mathsf{isAsync}\ #1}
\newcommand\isResidue[2]{\withmath{\texttt{isResidue}_{#1}\ #2}}
\newcommand\isSync[1]{\mathsf{isSync}\ #1}
\newcommand\noAsync[1]{\mathsf{noAsync}\ #1}
\newcommand\isLocal[1]{\mathsf{isLocal}\ #1}
\newcommand\isRemote[2]{\mathsf{isRemote}_{#1}\ #2}

\newcommand\finish[1]{\texttt{finish}\ #1}
\newcommand\LambdaFinish[1]{\withmath{\texttt{finish}_{\lambda}}\ #1}
\newcommand\FAILFinish[1]{\withmath{\texttt{finish}_{\FAIL}}\ #1}
\newcommand\MuFinish[1]{\withmath{\texttt{finish}_{\mu}}\ #1}
\newcommand\MuLambdaFinish[1]{\withmath{\texttt{finish}_{\mu+\lambda}}\ #1}

\newcommand\places{{\mathit places}\ }
\newcommand\Advance[0]{\texttt{advance}}
\newcommand\await[0]{\texttt{await}}

\newcommand\fail[0]{\texttt{fail}}
\newcommand\throw[1]{\texttt{throw}\,#1}
\newcommand\failed[0]{\texttt{failed}}

\newcommand\rd[2]{{\mathit rd}(#1,#2)}
\def\wr#1#2{{\mathit wr}(#1,#2)}
\newcommand\FOR[4]{\texttt{for}(#1\ \texttt{in}\ #2..#3)\ #4}
\def\O#1{{\mathcal O}\LL{#1}}

\def\P#1{{\mathcal P}\LL{#1}}
\newcommand\E[1]{{\mathcal E}\LL{#1}}

\newcommand\restr[2]{#1|_{#2}}

\def\from#1\infer#2{{{\textstyle #1}\over{\textstyle #2}}}
\def\rname#1\from#2\infer#3{{{\textstyle #2}\over{\textstyle #3}}{\ \mbox{#1}}}
\def\rnames#1\side#2\from#3\infer#4{{{\textstyle #3}\over{\textstyle #4}}{\ \textstyle(#2)}{\ \textstyle(#1)}}
\def\axname#1\axiom#2{{\textstyle #2}{\ \mbox{#1}}}
\def\WVFAILderives{\stackrel{\WVFAIL}{\derives}}
\def\VFAILderives{\stackrel{\VFAIL}{\derives}}
\def\DPFAILderives{\stackrel{\DPFAIL}{\derives}}
\def\Lambdaderives{\stackrel{\lambda}{\derives}}
\def\LambdaWderives{\stackrel{\lambda\WVFAIL}{\derives}}
\def\MuLambdaderives{\stackrel{\mu+\lambda}{\derives}}


\newcommand{\labeltree}[3]
{\begin{tabular}{l}
 \textsc{#1} \\[1mm]
 {\prooftree
 \text{$#2$}
 \justifies
 \raisebox{-1mm}{\text{$#3$}}
 \endprooftree}
 \end{tabular}
}

\newcommand{\labeltreeside}[4]
{\begin{tabular}{l}
 \textsc{#1} \\[1mm]
 {\infer[#4]{#3}{#2}}
 \end{tabular}
}

\newcommand{\olds}[1]{\oldstylenums{#1}}
\newcommand{\oldsb}[1]{{\bfseries\olds{#1}}}
\newcommand{\mnote}[1]{\stepcounter{ncomm}%
\vbox to0pt{\vss\llap{\scriptsize\oldsb{\arabic{ncomm}}}\vskip6pt}%
\marginpar{\scriptsize\bf\raggedright%
{\oldsb{\arabic{ncomm}}}.\hskip0.5em#1}}
\newcounter{ncomm}

\newcommand{\ignore}[1]{}


\renewcommand{\skip}{\texttt{skip}}
\newcommand{\dynat}[2]{\withmath{\overline{\texttt{at}}}\,(#1)\,#2}

\newcommand{\conf}[2]{\langle #1,#2 \rangle}
\newcommand{\reduce}{\longrightarrow}
\newcommand{\Reduce}{\Longrightarrow}

\newcommand{\isasync}[1]{\mathsf{isAsync}\ #1}
\newcommand{\issync}[1]{\mathsf{isSync}\ #1}
\renewcommand{\null}{{\mathit null}}
\newcommand{\spawned}[1]{\withmath{\overline{\texttt{async}}\,}#1}

\newcommand{\globalref}[1]{\texttt{globalref}\ #1}
\newcommand{\valof}[1]{\texttt{valof}\, #1}

\newcommand{\lred}[1]{\stackrel{#1}{\reduce}}
\newcommand{\lRed}[1]{\stackrel{#1}{\Reduce}}
\newcommand{\llred}[2]
{ \setbox0=\hbox{$\ {}^{#1}\ $}
  \setbox1=\hbox{$\longrightarrow_{#2}$}
  \loop\setbox1=\hbox{$-$\kern-0.3em\unhbox1}\ifdim\wd1<\wd0\repeat
  \hbox{$\ \ \mathop{\box1}\limits^{#1}\ \ $}
}
\newcommand{\epsred}{\stackrel{\epsilon}{\reduce}}
\newcommand{\lambdared}{\stackrel{\lambda}{\reduce}}
\newcommand{\lambdapred}{\stackrel{\lambda'}{\reduce}}
\newcommand{\sFailred}[1]{\stackrel{#1\otimes}{\reduce}}
\newcommand{\DPEFailred}{\stackrel{\DPE\times}{\reduce}}
\newcommand{\MskaFailred}{\llred{\mathsf{Msk}_{p,g}(v\times)}{p}}
\newcommand{\aFailred}{\stackrel{v\times}{\reduce}}
\newcommand{\failred}{\stackrel{v\times\,\mathit{or}\,v\otimes}{\reduce}}
\newcommand{\finishmu}[2]{\withmath{\texttt{finish}_{#1}}\ #2}
\newcommand{\try}[2]{\texttt{try}\,#1\,\texttt{catch}\,#2}


\newcommand{\prg}[1]  {\texttt{#1}}

%
%
%
%
%
%

\title{Semantics of (Resilient) X10}
\author{Silvia Crafa\inst{1} 
David Cunningham\inst{2} 
Vijay Saraswat\inst{3} 
Avraham Shinnar\inst{3} 
Olivier Tardieu\inst{3}}
\institute{
University of Padova,  Padova, IT \\ \email{crafa@math.unipd.it} 
\and Google, Inc \\ \email{sparkprime@gmail.com}
\and IBM TJ Watson Research Center \\ \email{\{vsaraswa,shinnar,tardieu\}@us.ibm.com}
}
\maketitle
\begin{abstract}
  We present a formal small-step structural operational semantics for
  a large fragment of X10, unifying past work. The fragment covers
  multiple places, mutable objects on the heap, sequencing,
  \code{try/catch}, \code{async}, \code{finish}, and \code{at}
  constructs.  This model accurately captures the behavior of a large
  class of concurrent, multi-place X10 programs. Further, we introduce
  a formal model of resilience in X10. During execution of an X10
  program, a place may fail for many reasons. Resilient X10 permits
  the program to continue executing, losing the data at the failed
  place, and most of the control state, and repairing the global
  control state in such a way that key semantic principles hold, the
  Invariant Happens Before Principle, and the Failure Masking
  Principle. These principles permit an X10 programmer to write clean
  code that continues to work in the presence of place failure.  The given
  semantics have additionally been mechanized in Coq.
\end{abstract}

\section{Introduction}
\label{sec:intro}

The need for scale-out programming languages is now well-established, 
because of high performance computing applications on supercomputers,
and analytic computations on big data.  Such languages -- based for
example on a partitioned global address space
(\cite{x10-concur05,x10-oopsla05}, \cite{upc2005upc}) -- permit programmers
to write a single program that runs on a collection of places on a
cluster of computers, can create global data-structures spanning
multiple places, can spawn tasks at remote places, detect termination
of an arbitrary tree of spawned tasks etc. The power of such languages
is shown by programs such as M3R, which implement a high-performance,
main-memory version of Hadoop Map Reduce \cite{Shinnar:2012:MIP:2367502.2367513} in a few thousand
lines of code. Other high performance multi-place libraries have been
developed for graph computations \cite{scale-graph} and sparse matrix 
computations \cite{gml}. 

At the same time, the practical realities of running large-scale
computations on clusters of commodity computers in commercial data
centers are that nodes may fail (or may be brought down, e.g. for
maintenance) during program executions. This is why multi-place
application frameworks such as
Hadoop \cite{dean-mr}, Resilient Data
Sets \cite{Zaharia:2010:SCC:1863103.1863113}, Pregel \cite{pregel} and
MillWheel \cite{millwheel} support resilient computations out of the
box. In case of node failure, relevant portions of the user computation are restarted. 

A new direction has been proposed recently in \cite{ResilientX10}:
extending a general purpose object-oriented, scale-out programming
language (X10) to support resilience. The hypothesis is that application
frameworks such as the ones discussed above can in fact be programmed
in a much simpler and more direct fashion in an object-oriented
language (powerful enough to build parallel, distributed libraries)
that already supports resilience.  
It is feasible to extend X10 in this way since is based on a few,
orthogonal constructs organized around the idea of {\em places} and
{\em asynchrony}. A place (typically realized as a process) is simply a
collection of objects together with the threads that operate on
them. A single computation may have tens of thousands of places. 
The statement \prg{async}~\prg{S} supports asynchronous execution
of \prg{S} in a separate task. \prg{finish}~\prg{S} executes \prg{S},
and waits for all tasks spawned by  \prg{S} to terminate.
Memory locations in one place can contain references ({\em global refs}) to
locations at other places. To use a global ref, the \prg{at}~\prg{(p)}~\prg{S} statement must
be used. It permits the current task to change its place of execution to
\prg{p}, execute \prg{S} at \prg{p} and return, leaving behind tasks that
may have been spawned during the execution of \prg{S}. The termination of
these tasks is detected by the \prg{finish} within which the \prg{at}
statement is executing. The values of variables used in \prg{S} but defined
outside \prg{S} are serialized, transmitted to \prg{p}, de-serialized to
reconstruct a binding environment in which \prg{S} is executed. Constructs are
provided for unconditional (\prg{atomic}~\prg{S}) and conditional (\prg{when}~\prg{(c)}~\prg{S})
atomic execution.  Finally, Java-style non-resumptive exceptions (\prg{throw},
\prg{try}/\prg{catch}) are supported.  If an exception is not caught in an
\prg{async}, it is propagated to the enclosing \prg{finish} statement.  Since
there may be many such exceptions, they appear wrapped in a
\prg{MultipleExceptions} exception.

\cite{ResilientX10} shows that this programming model may be extended
to support resilience in a surprisingly straightforward way. A
place \prg{p} may fail at any time with the loss of its heap and
tasks.  Any executing (or subsequent) tasks on that place
throw a \prg{DeadPlaceException} (\prg{DPE}).  Global refs pointing to
locations hosted at \prg{p} now ``dangle''; however they can only be
dereferenced via an \prg{at}~\prg{(p)}~\prg{S}, and this will throw
a \prg{DPE} exception.  If a task at a failed place has started
a task $T$ at another place, this task is not aborted. Instead
Resilient X10 posits a high level principle, the Happens Before
Invariance (HBI) principle: Failure of a place should not alter the happens
before relationship between statement instances at remaining
places. \cite{ResilientX10} shows that many interesting styles of
resilient programming can be expressed in Resilient X10. The
language is implemented at fairly modest cost.

In this paper we formalize the semantics of Resilient X10.  Our
fundamental motivation is to provide a  mechanized, formal semantics
for a core fragment of Resilient X10 that is separate from the
implementation and can be used as a basis for reasoning about
properties of programs and for establishing that principles such as
HBI actually hold.  

We proceed as follows. Our first task is to formalize a large portion
of X10. We build on the small-step, transition system for X10 presented
in \cite{yuki2013array} which deals with \prg{finish}, \prg{async}
and \prg{for} loops. We  extend it to handle multiple 
places and \code{at}, exceptions and try/catch statements, necessary
to express place failure. (In the spirit of \cite{yuki2013array} we omit
formalization of any of the object-oriented features of X10 since it
is fairly routine.) Configurations are just pairs $\tuple{s,g}$
representing a statement $s$ (the program to be executed) and a global
heap $g$, a partial map from the set of places to heaps.  Transitions
are (potentially) labeled with exceptions, tagged with whether they
were generated from a synchronous or asynchronous context. We
establish desirable properties of the transition system (absence of
stuck states, invariance of place-local heaps). We establish
a bisimulation based semantics that is consistent with the intuitions
underlying the ``gap based'' trace set semantics of
Brookes \cite{brookes-CSL}. We establish a set of equational laws for
this semantics.  

On this foundation we show that the semantics of Resilient X10 can be
formalized with just three kinds of changes. 
(1)~A place failure transition models the failure of a place $p$ by simply
removing $p$ from the domain of $g$. This cleanly models loss of all
data at $p$. Next, the transition rules for various
language constructs are modified to reflect what happens when those
constructs are ``executed'' at a failed place. (2)~An attempt to activate
any statement at a failed place results in
a \prg{DeadPlaceException} (abbreviated henceforth
as \prg{DPE}). (3)~Consistent with the design of Resilient X10, any
exceptions thrown by (the dynamic version of) an \prg{at(q) s} at a
failed place \prg{q} are masked by a \prg{DPE}. These are the only
changes needed. 

We show that the main properties of TX10 carry over to Resilient
TX10. We also show important resilience-related properties. Our main
theorem establishes that in fact Resilient TX10 satisfies Happens
Before Invariance. We also present a set of equational laws and
discuss differences with the laws for TX10.

We have encoded a mechanized version of the syntax and semantics of
both TX10 and Resilient X10 in Coq, an interactive theorem prover
\cite{coqart}. In doing so we addressed the challenge of formalizing the
copy operation on heaps and establishing termination (even in the
presence of cycles in the object graph). We mechanize the proof that
there are no stuck configurations, and furthermore prove that the
relation is computable, yielding a verified interpreter for TX10 and Resilient X10.

\paragraph{Related work.}
Our work is related to three broad streams of work. 
The first is formalization of X10 and Java with RMI. 
The first formalization of X10 was in \cite{x10-concur05}. This paper
adapts the framework of Middleweight Java \cite{MJ}
to represent a configuration as a collection of stacks and heaps. This
choice led to a rather complex formalization. \cite{lee2010feather}
presents an operational semantics for the X10 \code{finish}/\code{async} fragment,
but again with a complex representation of control. We build on the
work of \cite{yuki2013array} which for the first time represents the
control state as a statement, and presents a very simple definition of
the Happens Before relation. We extend that work to handle exceptions
(necessary for the formalization of resilience), and place-shifting
\code{at}, and formally treat resilience. \cite{rmi-oopsla05} presents
a semantics for Java with remote method invocation; hence they also
deal with multiple places and communication across places. In
particular they formalize a relational definition of copying an object
graph, although they do not formalize or mechanize an implementation
of this specification. Their formalization does not deal with
place failure, since Java RMI does not deal with it.

The second stream is the work on formalization of the
semantics of concurrent imperative
languages \cite{Brookes93fullabstraction,failureOfFailures,brookes-CSL}. Our work can 
be seen as adding block-structured concurrency
constructs (\prg{finish}, \prg{async}), exceptions, and, of course,
dealing with  multiple places, and place failure. 

The third stream is the work on distributed process algebras that deal with
failure \cite{Fournet96acalculus,DPI,DBLP:journals/iandc/FrancalanzaH08,amadio-failure,Riely:2001:DPL:504563.504586}. 
\cite{amadio-failure} introduces an extension of the $\pi$-calculus
with located actions, in the context of a higher-order, distributed
programming language, Facile. 
\cite{Fournet96acalculus} introduces locations in the distributed join
calculus, mobility and the possibility of location failure, similar to
our place failure. The failure of a location can be detected, allowing
failure recovery. In the context of D$\pi$ \cite{DPI}, an extension of the $\pi$-calculus with
multiple places and mobility, \cite{DBLP:journals/iandc/FrancalanzaH08} gives a treatment
of node- and link-failure. In relationship with all these works, this
work differs in dealing with resilience in the context of distributed state, global references, mobile
tasks with distributed termination detection (finish), and
exceptions, and formalizing the HBI principle. Our work is motivated
by formalizing a real resilient programming language, rather than working with abstract
calculii.

\paragraph{Summary of Contributions.} The contributions of this paper are:
\begin{itemize}
\item We present a formal operational semantics for significant
  fragment of X10, including multiple places, mutable heap,
  \code{try/catch} statements, \code{throws}, \code{async},
  \code{finish} and \code{at} 
  statements. The semantics is defined in terms of a labeled
  transition relation over configurations in which the control state
  is represented merely as a statement, and the data state as a
  mapping from places to heaps. 
\item We present a set of equational laws for operational congruence.
\item We extend the formal operational semantics to Resilient X10,
 showing that it enjoys Happens Before Invariance and Failure Masking
 Principles. 
\item We present equational laws for Resilient X10.
\item We mechanize proofs of various propositions in Coq. In
 particular, the proof that no configurations are stuck yields a verified
 executable version of the semantics.
\end{itemize}



\paragraph{Rest of this paper.}
Section~\ref{sec:tx10} introduces TX10, informally describing the basic constructs
and a small-step operational semantics of TX10.
Section~\ref{sec:equational-laws} 
presents laws for equality for a
semantics built on congruence over bisimulation. The second half of
the paper presents a semantic treatment of
resilience. Section~\ref{sec:resilientTX10} 
discusses the design of Resilient X10, formalizes the semantics,
and presents equational laws for congruence. Section~\ref{sec:conclusion} concludes.

\section{TX10}
\label{sec:tx10}

We describe in this section the syntax and the semantics of TX10, the
formal subset of the X10 language \cite{X10} we consider in this work.
We have also encoded a mechanized version in Coq, which will be
discussed in Section~\ref{subsec:x10-mechanization-coq}.


\begin{table}[t]
\begin{center}
\begin{tabular}{ll}
\begin{tabular}{l}
\mbox{
\begin{tabular}{lllll}
\multicolumn{2}{l}{(Values)} & $v,w$ & {::=} & \\
&\multicolumn{2}{l}{$o$} &\multicolumn{2}{l}{\mbox{\em (Runtime only.)
    Object ids}}\\
&\multicolumn{2}{l}{$o\$p$} &\multicolumn{2}{l}{\mbox{\em (Runtime
    only.) Global Object ids}}\\
&\multicolumn{2}{l}{${\mathsf E},\BFE,\BGE,\DPE$} &\multicolumn{2}{l}{\mbox{\em Exceptions}}\\
\quad\\
\end{tabular}
}
\\
\mbox{
\begin{tabular}{lllll}
\multicolumn{2}{l}{(Programs)} & $pr$ & {::=} & \\
&\multicolumn{2}{l}{\finish{\Atttwo{0}{$s$}}}&\multicolumn{2}{l}{\em
  activation} \\
\end{tabular}
}
\end{tabular}
&
\mbox{
\begin{tabular}{lllll}
\multicolumn{2}{l}{(Expressions)} & $d,e$ & {::=} & \\
&\multicolumn{2}{l}{$v$} &\multicolumn{2}{l}{\mbox{\em Values}}\\
&\multicolumn{2}{l}{\texttt{x}} &\multicolumn{2}{l}{\mbox{\em Variable access}}\\
&\multicolumn{2}{l}{$e.f$} &\multicolumn{2}{l}{\mbox{\em Field selection}}\\
&\multicolumn{2}{l}{$\{f{:}e,\ldots, f{:}e$\}} &\multicolumn{2}{l}{\mbox{\em Object construction}}\\
&\multicolumn{2}{l}{\code{globalref} $e$}   &\multicolumn{2}{l}{\mbox{\em GlobalRef construction}}\\
&\multicolumn{2}{l}{\code{valof} $e$}       &\multicolumn{2}{l}{\mbox{\em Global ref deconstruction}}\\
\quad\\
\end{tabular}
}
\end{tabular}

\begin{tabular}{c}
\mbox{
\begin{tabular}{lllll}
\multicolumn{2}{l}{(Statements)} & $s,t$ & {::=} & \\
&\multicolumn{2}{l}{\texttt{skip;}} &\multicolumn{2}{l}{\mbox{\em Skip -- do nothing}}\\
&\multicolumn{2}{l}{\throw{v};}&\multicolumn{2}{l}{\em Throw an exception} \\
&\multicolumn{2}{l}{$\ValS{x}{e}{s}$} &\multicolumn{2}{l}{\mbox{\em Let bind e to x in s}}\\
&\multicolumn{2}{l}{$e.f=e;$} &\multicolumn{2}{l}{\mbox{\em  Assign to field}}\\
&\multicolumn{2}{l}{$\{s\ t\}$} &\multicolumn{2}{l}{\mbox{\em Run $s$ then $t$}}\\
&\multicolumn{2}{l}{$\Att{p}{x}{e}{s}$}&\multicolumn{2}{l}{\em Run $s$ at $p$ with $x$ bound to $e$} \\
&\multicolumn{2}{l}{$\async{s}$}&\multicolumn{2}{l}{\mbox{\em Spawn $s$ in a different task}}\\
&\multicolumn{2}{l}{$\finish{s}$} &\multicolumn{2}{l}{\mbox{\em Run $s$ and wait for termination}}\\
&\multicolumn{2}{l}{$\TryCatch{s}{t}$} &\multicolumn{2}{l}{\mbox{\em Try $s$, on failure execute $t$}}\\
&\multicolumn{2}{l}{$z$} &\multicolumn{2}{l}{\mbox{\em Runtime versions}}\\
\end{tabular}
}
\end{tabular}
\\
\mbox{
\begin{tabular}{lllll}
\multicolumn{2}{l}{(Dynamic Stmts)} & $z$ & {::=} & \\
&\multicolumn{2}{l}{$\At{p}{s}$}&\multicolumn{2}{l}{\em Runtime only} \\
&\multicolumn{2}{l}{$\asynca{s}$}&\multicolumn{2}{l}{\mbox{\em Runtime only}}\\
&\multicolumn{2}{l}{$\MuFinish{s}$} &\multicolumn{2}{l}{\mbox{\em Run $s$, recording exceptions in $\mu$}}
\\
\end{tabular}
}
\end{center}
\caption{Syntax of TX10}\label{table:tx10}
\end{table}


The syntax of TX10 is defined in Table~\ref{table:tx10}. 
We assume an infinite set of values \Val, ranged over by $v,w$,
an infinite set of variables ranged over by {\texttt x,y}, and an
infinite set of field names ranged over by $f$.
We also let $p,q$ range over a finite set of integers 
\Places$= 0...(n{-}1)$, which represent available computation
\emph{places}.  
%
A source program is defined as a static statement $s$ activated at
place $0$ under a governing \code{finish} construct. The syntax 
then includes dynamic statements and dynamic values that can only
appear at runtime. 
Programs operate over objects, either local or global, that are
handled through object identifiers (object ids). We assume an infinite
set of object ids, \Loc{} (with a given bijection with the natural
numbers, the ``enumeration order''); objects are in a one to one
correspondence with object ids. Given the distributed nature of the
language and to model X10's \code{GlobalRef}, we assume that each
object lives in a specific (home) place, and we distinguish between
local and global references. More precisely, we use the
following notation: 
\begin{itemize}
\item $\mathsf{p}:\Loc\to \Places$ maps each object id to the place
  where it lives;  
\item $\Loc_q=\{o\in \Loc~|~ \mathsf{p}(o)=q\}$ and
     $\gLoc=\{o\$ p ~|~ o\in\Loc_p \ \wedge \ p\in\Places\}$
\end{itemize}
Then given $o\in\Loc_q$, we say that $o$ is a local reference (to a
local object) while $o\$q$ is a global reference (to an object located
at $q$).

The expression $\{f_1:e_1, \ldots, f_n:e_n\}$ (for $n \geq 0$) creates
a new local object and returns its fresh id. The object is initialized
by setting, in turn, the fields $f_i$ to the value obtained by
evaluating $e_i$. 
Local objects support field selection: the expression
$e.f$ evaluates to the value of the field with name $f$ in the object 
whose id is obtained by evaluating $e$. Similarly, the syntax of
statements allows field update.  
X10 relies on a type system to ensure that any selection/update
operation occurring at runtime is performed on an object that actually
contains the selected/updated field. Since TX10 has no corresponding
static semantic rules, we shall specify that $o.f$ throws a
\emph{BadFieldSelection} $\BFE$ exception when the object $o$ does not
have field $f$.  

The expression \code{globalref e} creates a new global
reference for the reference returned by the evaluation of
\code{e}. Whenever \code{e} evaluates to a global reference, the
expression \code{valof e} returns the local object pointed by
\code{e}. Errors in dealing with global references are modelled by
throwing a \emph{BadGlobalRef} exception $\BGE$.
(see Section~\ref{sec:tx10-semantics} for a detailed explanation of
the semantics of global references).

TX10 deals with exception handling in a standard way: the statement
$\throw{v}$ throws an exception value $v$ that can be caught with a
$\TryCatch{s}{t}$ statement. For simplicity, exception values are
constants: besides $\BFE$ and $\BGE$ described above, we add 
${\mathsf E}$ to represent a generic exception. The exception $\DPE$
stands for \emph{DeadPlaceException}, and will only appear in the
semantics of the resilient calculus in
Section~\ref{sec:resilientTX10}. 
Variable declaration $\ValS{x}{e}{s}$ declares a new local variable
$x$, binds it to the value of the expression $e$ and continues as
$s$. The value assigned to $x$ cannot be changed during the
computation. We shall assume that the only free variable of $s$ is $x$
and that $s$ does not contain a sub-statement that declares the same
variable $x$.
This statement is a variant of
the variable declaration available in X10. In X10 the scope $s$ is not
marked explicitly; rather all statements in the rest of the current
block  are in scope of the declaration. We have chosen this ``let''
variant to simplify the formal presentation.

The construct $\async{s}$ spawns an independent lightweight thread,
called {\em activity}, to execute $s$. The new activity running in
parallel is represented by the dynamic statement $\spawned{s}$. 
The statement $\finish{s}$ executes $s$ and waits for the termination
of all the activities (recursively) spawned during this execution.
Activities may terminate either normally or abruptly, {\em i.e.} by
throwing an exception. If one or more activities terminated abruptly, 
$\finish{s}$ will itself throw an exception that encapsulates all
exceptions. In TX10, we use the parameter $\mu$ in \MuFinish{s} to
record the exception values thrown by activities in $s$. $\mu$ is a
possibly empty set of values; we simply write $\finish{s}$
instead of $\finishmu{\emptyset}{s}$.


The sequence statement $\{s\,t\}$ executes $t$ after executing
$s$. Note that if $s$ is an \code{async}, its execution will simply
spawn an activity $\spawned{s}$, and then activates $t$. Therefore, 
$\{\spawned{s}\ \ t\}$ will actually represent $s$ and $t$ executing
in parallel. We say that sequencing in X10 has {\em shallow
  \code{finish} semantics}

Finally, $\Att{p}{x}{e}{s}$ is the place-shifting statement. We
assume that the only free variable in $s$ is $x$. This statement first
evaluates $e$ to a value $v$, then copies the object graph rooted at
$v$ to place $p$ to obtain a value $v'$, and finally executes $s$
synchronously at $p$ with $x$  bound to $v'$. 
Running $s$ at $p$ synchronously means that in
$\{\Att{p}{x}{e}{s}\ \  t\}$, $t$ will be enabled precisely when
the \texttt{at} statement has only asynchronous sub-statements left
(if any).  Thus \texttt{at} also has shallow \code{finish} semantics, just
like sequential composition. 
In some cases the programmer may not need to transmit values from
the calling environment to $s$, the variant $\Atttwo{p}{s}$ may be
used instead. 
As an example, the program 
$\finish{\at{0}{\{\at{1}{\async{s}}\ \ \ \at{2}{\async{s}}\}}}$
evolves to a state where two copies of $s$ run in parallel at places
$1$ and $2$. The entire program terminates whenever both remote
computations end. 

Currently, X10 supports a variant of these \code{at} constructs.
The programmer writes $\Atttwo{p}{s}$ and the
compiler figures out the set of variables used in $s$ and declared
outside $s$. A copy is made of the object reference graph with the
values of these variables as roots, and $s$ is executed with these
roots bound to this copied graph.  
Moreover X10, of course, permits mutually recursive procedure (method)
definitions. We leave the treatment of recursion as future work.


\subsection{Operational Semantics}\label{sec:tx10-semantics}

We build on the semantics for X10 presented in \cite{yuki2013array}.
In this semantics, 
the data state is maintained in a shared global heap (one heap per
place), but the control state is represented in a block structured
manner -- it is simply a statement. 
$$
\begin{array}{l}
\mathit{Heap}\ \ h\ ::= \ \emptyset ~|~ h\cdot[o\mapsto r]
\quad\quad
\mathit{Global\ heap}\ \ g\ ::= \ \emptyset ~|~ g\cdot[p\mapsto h]
\end{array}
$$
The local heap at a place $p$ is a partial map that associates object
ids to objects represented by partial maps $r$ from field names to
object ids. The global heap $g$ is a partial map 
form the set of places $\Places$ to local heaps.
We let $\emptyset$ denote the unique partial map with empty domain,
and for any partial map $f$ by $f[p\to v]$ we mean the map $f'$ that
is the same as $f$ except that it takes on the value $v$ at $p$. 
Moreover, in the following we write $s[^v/_x]$ for variable
substitution. 

X10 is designed so that at run-time heaps satisfy the 
{\em place-locality invariant} formalized below. Intuitively, 
the domain of any local heap only contains local object references,
moreover any object graph (rooted at a local object) only contains 
references to either (well defined) local objects or global
references.

Let $h$ be a local heap and $o\in\dom{h}$ an object identifier.
We let $h{\downarrow}_o$ denote {\em the object graph rooted at $o$},
that is the graph with vertexes the values reachable from $o$ via the
fields of $o$ or of one or more intermediaries. In other terms, it is
the graph where an $f$-labelled edge $(v,f,v')$ connects the vertices's
$v$ and $v'$ whenever $v$ is an object with a field $f$ whose value is
$v'$. We also denote by $h_o$ the set of all object values that are
reachable from $o$, that is the set of all vertices's in the object
graph $h{\downarrow}_o$. 

\begin{definition}[Place-local heap]
A global heap $g$ is {\em place-local} whenever 
for every $q\in \dom{g}$, and $h=g(q)$
\begin{itemize}
\item $\dom{h} \subseteq \Loc_q$ and
$\forall o\in \dom{h}. \ \ h_{o} \subseteq (\Loc_q\cap\dom{h})
  \cup \gLoc$
\end{itemize}
\end{definition}


%

\noindent
The semantics is given in terms of a transition relation between 
{\em configurations}, which are
either a pair $\tuple{s,g}$ (representing the statement $s$ to be
executed in global heap $g$) or a singleton $g$, representing a
computation that has terminated in $g$. Let $k$ range over
configurations. 
The transition relation $k\Lambdaderives_pk'$ is defined as a labeled
binary relation on configurations, where  
$\lambda \in \Lambda=\{\epsilon,\WVFAIL,\VFAIL\}$, and $p$
ranges over the set of places. 
The transition $k \Lambdaderives_p k'$ is to be understood as: the
configuration $k$ executing at $p$ can in one step evolve to $k'$,
with $\lambda=\epsilon$ indicating a normal transition, and 
$\lambda=\VFAIL$, resp. $\WVFAIL$, indicating that an exception has
thrown a value $v$ in a synchronous, resp. asynchronous, subcontext. 
Note that failure is not fatal; a failed transition may be
followed by any number of failed or normal transitions. We shall write
$\stackrel{\epsilon}{\derives}_p$ as $\derives_p$.

\begin{definition}[Semantics]
Let $\starderives$ represent the reflexive, transitive closure of
$\Lambdaderives_{0}$. 
The operational semantics, $\O{s}$ of a statement $s$ is the relation
$$
\O{s}\defeq \{(g,g') \alt \tuple{\finish\At{0}{s},g}\starderives g'\}
$$
\end{definition}

\noindent
In order to present rules compactly, we use the ``matrix'' convention
exemplified below, 
where we write the left-most rule to compactly denote the four rules
obtained from the right-most rule with $i=0,1, j=0,1$. 
{\small
$$
\begin{array}{ccc}
\infer{
\begin{array}{ll} 
cond_0\ &  \delta^0 \lred{\lambda_0}\delta^0_{0}\alt \delta^0_{1} \\[1mm] 
cond_1\ &  \delta^1 \lred{\lambda_1} \delta^1_{0} \alt \delta^1_{1}
   \end{array}
}
   { \gamma \lred{\lambda} \gamma_0 \alt\gamma_{1} }
\quad\quad\quad
&
\quad\quad\quad
\infer{\delta^j \lred{\lambda_j} \delta^j_i }
   { \gamma \lred{\lambda} \gamma_i  \quad cond_j}
&
\quad i=0,1\ j=0,1
\end{array}
$$
}

%

\begin{table}[t]
$$
\begin{array}{c}
\vdash\isasync{\spawned{s}} 
\quad\quad\quad
\infer{\begin{array}{l}
        \vdash\isasync{\dynat{p}{s}}\\
        \vdash\isasync{\try{s}{t}}
       \end{array}
       }{\vdash\isasync{s}}
\quad\quad\quad
\infer{\vdash\isasync{\{s\ t\}}}{\vdash\isasync{s}& \vdash\isasync{t}}
\\ \\
\begin{array}{l}
\vdash\issync{s^*} \\ 
 \mbox{ with }
     s^*\in \left\{ \begin{array}{l}
               \skip,\ \texttt{val x=e s},\ e.f=e, \\
               \Att{p}{x}{e}{s},\ \async{s},\\
               \finish_\mu{\ s}, \throw{v} 
               \end{array}
               \right\}
\end{array}
\quad\quad\quad
\infer{\begin{array}{l}
       \vdash\issync{\{s\ t\}}\\
        \vdash\issync{\{t\ s\}}\\
        \vdash\issync{\dynat{p}{s}}\\
         \vdash\issync{\try{s}{t}}
        \end{array}
}{\vdash\issync{s}}
\\ \\
\end{array}
$$
\caption{Synchronous and Asynchronous Statements}
\label{tab:auxiliary}
\end{table}

\noindent 
We also introduce in Table~\ref{tab:auxiliary} two auxiliary
predicates to distinguish between {\em asynchronous} and {\em
  synchronous} statements. 
A statement is asynchronous if it is an $\spawned{s}$, or a sequential
composition of asynchronous statements (possibly running at other
places).  
The following proposition is easily established by structural
induction. 
\begin{proposition}
For any statement $s$, either \mbox{$\vdash \isAsync{s}$} xor
\mbox{$\vdash \isSync{s}$}. 
\end{proposition}

\begin{table}[Ht]
{\small
\begin{center}
\begin{tabular}{c}
\labeltree{(New Obj)}
{o\in \lLoc_p{\setminus}\dom{h} \quad \quad n \geq 0}
{\conf{\{f_1{:}v_1,...,f_n{:}v_n\}}{h}\reduce_p
 \conf{o}{h\cdot[o\mapsto \emptyset[f_1 \mapsto v_1] \ldots [f_n \mapsto v_n]]} }
\\ \\
\labeltree{(Select)}
{h(o){=} r[f \mapsto v]\}}
{\conf{o.f}{h}\reduce_p\conf{v}{h}}
\quad
\labeltree{(Select Bad)}
{
 f{\notin}\dom{h(o)}}
{\conf{o.f}{h}\sFailred{\BFE}_p h}
\quad
\labeltree{(New Global Ref)}
{}
{\conf{\globalref{o}}{h}\reduce_p\conf{o\$ p}{h}}
\\ \\ 
\labeltree{(Valof)}
{}
{\conf{\valof{o\$ p}}{h}\reduce_p\conf{o}{h}}
\quad\quad 
\labeltree{(Valof Bad)}
{v\neq o\$ p}
{\conf{\valof{v}}{h}\sFailred{\BGE}_p h}
\\ \\
\labeltree{(Exp Ctx)}
{\conf{e}{h}\lambdared_p \conf{e'}{h'}~|~h}
{\begin{array}{rcl}
\conf{e.f}{h}& \lambdared_p & \conf{e'.f}{h'}~|~h\\
\conf{\globalref{e}}{h}& \lambdared_p & \conf{\globalref{e'}}{h'}~|~h\\
\conf{\valof{e}}{h}& \lambdared_p & \conf{\valof{e'}}{h'}~|~h\\
\conf{\{f_1{:}v_1,...,f_{i}{:}v_{i},f_{i+1}{:}e,...\}}{h}
    & \lambdared_p &
   \conf{\{f_1{:}v_1,...,f_{i}{:}v_{i},f_{i+1}{:}e',...\}}{h'}~|~h
\end{array}
}
\end{tabular}
\end{center}
}
\caption{Expression Evaluation}
\label{tab:ExprEval}
\end{table}

In order to define the transition between configurations,
we first define the evaluation relation for expressions by the rules
in Table~\ref{tab:ExprEval}. 
%
Transitions of the form $\conf{e}{h}\reduce_p\conf{e'}{h'}$ state that
the expression $e$ at place $p$ with local heap $h$ correctly
evaluates to $e'$ with heap $h'$. On the other hand an error in the
evaluation of $e$ is modeled by the transition
$\conf{e}{h}\sFailred{v}_p h$. 
%
An object creation expression is evaluated from left to right,
according to rule {\sc (Exp Ctx)}. When all expressions are evaluated,
rule {\sc (New Obj)} states that a new local object id is created and
its fields set appropriately. Rule {\sc (New Global Ref)} shows that a
new global reference is built from an object id $o$ by means of
the expression $\globalref{o}$. 
A global reference $o\$ p$ can be dereferenced by means of the
\code{valof} expression. Notice that rule {\sc (Valof)}, according to
X10's semantics, shows that the actual object can only be accessed
form its home place, i.e. $\mathsf{p}(o)=p$. 
Any attempt to select a non-existing field from an object results in
the $\BFE$ exception by rule {\sc (Select Bad)}, while any attempt to
access a global object that is not locally defined result in a $\BGE$
error by rule {\sc (Valof Bad)}. 
In X10, the static semantics guarantees that objects and global
references are correctly created and that any attempt to select a
filed is type safe, hence well typed X10 programs do not occur in 
$\BFE$ and $\BGE$ exceptions, however we introduce rules 
{\sc (Select Bad), (Valof Bad)} and {\sc (Bad Field Update)} 
so that the operational
semantics of TX10 enjoys the property that there are no stuck states,
i.e. Proposition~\ref{prop:no-stuck} in
Section~\ref{subsec:properties}.

The following proposition shows that the heap modifications performed
by rules {\sc (New Obj)} and {\sc (New Global Ref)} respect
the place-locality invariant.

\begin{proposition}\label{prop:PlaceLocalExpr}
Let $g$ be a place-local heap, $p\in \dom{g}$ and $h=g(p)$. 
We say that $\conf{e}{h}$ is place-local whenever
for any local object id $o$ occurring in $e$ it
holds $o\in\dom{h}$. 
If $\conf{e}{h}$ is place-local and
$\conf{e}{h}\reduce_p\conf{e'}{h'}$, then
$g\cdot [p\mapsto h']$ is place-local, and
$\conf{e'}{h'}$ is place-local.
\end{proposition}


\begin{table}[t]
{\small
\begin{center}
\begin{tabular}{c}
\labeltree{(Skip)}
{p\in dom(g)}
{\conf{\skip}{g}\reduce_p g}
\quad
\labeltree{(Exception)}
{p\in dom(g)}
{\conf{\throw{v}}{g}\VFAILderives_p g} 
\quad
\labeltree{(Declare Val)}
{p\in dom(g)\quad \conf{s[^v/_x]}{g}\lambdared_p\conf{s'}{g'}~|~g'}
{\conf{\ValS{x}{v}{s}}{g}\lambdared_p\conf{s'}{g'}~|~g'} 
\\ \\
\labeltree{(Field Update)}
{p\in dom(g) \quad f\in \dom{g(p)(o)} }
{\conf{o.f=v}{g}\reduce_p 
   g[p\to g(p)[o\to g(p)(o)[f\mapsto v]]]}
\quad
\labeltree{(Bad Field Update)}
{p\in dom(g) \quad f\notin\dom{g(p)(o)} }
{\conf{o.f=v}{g}\sFailred{\BFE}_p g}
\\ \\
\labeltree{(Ctx)}
{p\in dom(g)\quad \conf{e}{g(p)}\lambdared_p\conf{e'}{h'}~|~h' \quad g'=g[p\mapsto h']}
{\begin{array}{rcl}
\conf{\ValS{x}{e}{s}}{g}
  & \lambdared_p & \conf{\ValS{x}{e'}{s}}{\ g'}~|~g' \\
\conf{e.f=e_1}{g}& \lambdared_p &\conf{e'.f=e_1}{\ g']}~|~g'\\
\conf{o.f=e}{g} & \lambdared_p & \conf{o.f=e'}{\ g'}~|~g'\\
\conf{\Att{p}{x}{e}{s}}{g} & \lambdared_p &
  \conf{\Att{p}{x}{e'}{s}}{\ g'}~|~g'
\end{array}
}
\end{tabular}
\end{center}
}
\caption{Basic Statements}
\label{tab:semBasic}
\end{table}

\noindent
Now we turn to the axiomatization of the transition relation
between configurations. 

%
Table~\ref{tab:semBasic} collects a first set of rules dealing with
basic statements. These rules use the condition $p\in\dom{g}$, which
is always true when places do not fail. We include this condition to
permit the rules of Table~\ref{tab:semBasic} to be reused when we
consider place failure in Section~\ref{sec:resilientTX10}.
Most of these rules are straightforward. 
Rule {\sc (Exception)} shows that throwing an exception is recorded as
a synchronous failure. Moreover, rule {\sc (Bad Field Update)} throws
a $\BFE$ exception whenever  $f$ is not one of its fields. 

\begin{table}[t]
{\small
\begin{center}
\begin{tabular}{c}
\labeltree{(Spawn)}
{}
{\conf{\async{s}}{g} \reduce_p \conf{\spawned{s}}{g} }
\quad
\labeltree{(Async)}
{\conf{s}{g}\lambdared_p\conf{s'}{g'}~|~g'}
{\begin{array}{ll}
\lambda{=}\epsilon & 
     \conf{\spawned{s}}{g} \reduce_p \conf{\spawned{s'}}{g'}~|~g' 
\\[1mm]
\lambda{=}v\times,v\otimes\ \ &
    \conf{\spawned{s}}{g} \llred{{\mathsf{Msk}(v\times)}}{p} \conf{\spawned{s'}}{g'}~|~g'
\end{array}
}
\\ \\
\labeltree{(Finish)}
{\conf{s}{g}\lambdared_p\conf{s'}{g'}}
{\conf{\finishmu{\mu}{\!s}}{g} \reduce_p  \conf{\finishmu{\mu\cup\lambda}{\!s'}}{g'}}
\quad\quad
\labeltree{(End of Finish)}
{\conf{s}{g}\lambdared_p g'
\quad\quad \lambda'{=}\left\{\begin{array}{ll}
              \epsilon & \mbox{ if } \lambda{\cup}\mu{=}\emptyset\\
              {\mathsf E}\otimes & \mbox{ if }
                    \lambda{\cup}\mu{\neq}\emptyset
              \end{array}
              \right.
}
{\conf{\finishmu{\mu}{s}}{g}\llred{{\mathsf{Msk}(\lambda')}}{p} g'}
\\ \\
%
%
\labeltree{(Seq)}
{ \conf{s}{g}\lambdared_p\conf{s'}{g'} ~|~ g'}
{\begin{array}{ll}
\lambda=\epsilon,v\times & \conf{\{s\ t\}}{g} \lambdared_p
        \conf{\{s'\ t\}}{g'}~|~ \conf{t}{g'}
\\[1mm]
\lambda=v\otimes & \conf{\{s\ t\}}{g} \lambdared_p
        \conf{s'}{g'}~|~ g'
 \end{array}
}
\quad\quad
\labeltree{(Par)}
{\vdash\isasync{t}\quad \conf{s}{g}\lambdared_p\conf{s'}{g'}~|~ g'}
{\conf{\{t\ s\}}{g} \lambdared_p
        \conf{\{t\ s'\}}{g'}~|~\conf{t}{g'}}
\\ \\
%
\labeltree{(Place Shift)}
{
 (v',g')= \mathsf{copy}(v,q,g)}
{\conf{\Att{q}{x\!}{\!v}{s}}{g} \reduce_p
  \conf{\dynat{q}{\{s[^{v'}\!/_x]\ \, \skip\}}}{g'}
}
\quad
\labeltree{(At)}
{
 \conf{s}{g}\lambdared_q\conf{s'}{g'}~|~g'}
{ \conf{\dynat{q}{s}}{g} \lambdared_p  \conf{\dynat{q}{s'}}{g'}~|~g'}
\\ \\
\labeltree{(Try)}
{
    \conf{s}{g}\lambdared_p\conf{s'}{g'}~|~ g'}
{\begin{array}{ll}
\lambda=\epsilon,v\times & 
    \conf{\try{s}{t}}{g} \lambdared_p  \conf{\try{s'}{t}}{g'}~|~g'
\\[1mm]
\lambda=v\otimes & 
    \conf{\try{s}{t}}{g} \reduce_p  \conf{\{s'\  t\}}{g'}~|~\conf{t}{g'}
\end{array}     
}
\end{tabular}
\end{center}
}
\caption{Statements Semantics}
\label{tab:semII}
\end{table}

The rest of operational rules are collected in Table~\ref{tab:semII}.
These rules, besides defining the behavior of the major
X10 constructs, also illustrate how the exceptions are propagated
through the system and possibly caught. 
The \code{async} construct takes one step to spawn the new
activity. Moreover, according to rule {\sc (Async)}, an exception
(either synchronous or asynchronous) 
in the execution of $s$ is {\em masked} by an asynchronous exception
in $\spawned{s}$. Asynchronous failures are confined within the thread
where they originated, and they are caught by the closest
\code{finish} construct that is waiting for the termination of such a
thread. 
More precisely,
the $\finish{s}$ statement waits for the termination of any (possibly
remote) asynchronous (and synchronous as well) activities spawned by
$s$. Any exception thrown during the evaluation of $s$ is absorbed and
recorded into the state of the governing \code{finish}. Indeed,
consider rule {\sc (Finish)} where we let be $\mu\cup\lambda{=}\mu$
if $\lambda{=}\epsilon$ and $\mu\cup\lambda{=}\{v\}\cup\mu$ if
$\lambda{=}v\times$ or $\lambda{=}v\otimes$.
Then this rule
shows that the consequence has a correct transition
$\reduce_p$ even when $\lambda\neq\epsilon$: i.e., the exception
in $s$ has been absorbed and recorded into the state of \code{finish}.
Moreover, the rule {\sc (End of Finish)} shows that \code{finish}
terminates with a generic synchronous exception whenever at least one
of the activities its governs threw an exception (in X10 it throws
a \code{MutipleExceptions} containing the list of exceptions collected
by \code{finish}).
Two rules describe the semantics of sequential composition.
When executing $\{s\ t\}$, rule {\sc (Seq)} shows that the
continuation $t$ is activated whenever $s$ terminates normally or with
an asynchronous exception. On the other hand, when the execution of
$s$ throws a synchronous exception (possibly leaving behind residual
statements $s'$) the continuation $t$ is discarded.
Rule {\sc (Par)} captures the essence of asynchronous
execution allowing reductions to occur in parallel
components.

The rule {\sc (Place Shift)} activates a remote computation; it 
uses a {\em copy} operation on object graphs, $\Copy{o}{q}{g}$, that
creates at place $q$ a copy of the object graph rooted at $o$,
respecting global references. In X10 place shift is implemented by
recursively serializing the object reference graph $G$ rooted at $o$
into a byte array. In this process, when it is encountered a global
object reference $o\$ p$, the fields of this object are not
followed; instead the unique identifier $o\$ p$ is serialized. The
byte array is then transported to $q$, and de-serialized at $q$ to
create a copy $G'$ of $G$ with root object a fresh identifier $o'\in
\Loc_q$.  All the objects in $G'$ are new. 
$G'$ is isomorphic to $G$ and has the additional property that if
$z$ is a global ref that is reachable from $o$ then it is also
reachable (through the same path) from $o'$. 

\begin{definition}[The copy operation.]
\label{def:copy}
Let $g$ be a global heap, $q$ a place with $h=g(q)$. 
Let be $o\in\Loc$ such that $\mathsf{p}(o)\in \dom{g}$,
then $\Copy{o}{q}{g}$ stands for the (unique) tuple
\tuple{o',g[q\rightarrow h']} satisfying the following properties,
where $N=\dom{h'}\setminus \dom{h}$.  
\begin{itemize}
\item $N$ is the next $|N|$ elements of $\Loc_q$.
\item $o'\in N$  
\item There is an isomorphism $\iota$ between the object graph
  $g(\mathsf{p}(o)){\downarrow}_o$ rooted at $o$ and the object graph 
  $h'{\downarrow}_{o'}$ rooted at $o'$. 
  Further, $\iota(v)=v$ for $v\in \gLoc$ 
\item $h'_{o'} \subseteq N\cup \gLoc$.
\item $h'=h\cdot[o'\mapsto r]$ where $r$ is the root object of the
  graph $h'{\downarrow}_{o'}$  
\end{itemize}
We extend this definition to arbitrary values, that is 
$\Copy{v}{q}{g}$ is defined to be $v$ unless $v$ is an object id, in
which case it is defined as above.
\end{definition}

\begin{proposition}\label{prop:CopyCorrect}
Let $g$ is place-local heap.
 Let $p,q\in \dom{g}$ be two (not necessarily distinct)
places, and let $o\in \Loc_p$. Let
$\Copy{o}{q}{g}=\tuple{o',g'}$. Then $g'$ is place-local.
\end{proposition}

Place-shift takes a step to activate. Moreover, in the
conclusion of the rule {\sc (Place Shift)}  
the target statement contains a final $\skip$ in order to model
the fact that the remote control has to come back at the local 
place after executing the remote code $s[^{v'}/_{x'}]$. This 
additional step is actually needed in the resilient calculus,
where we need to model the case where the remote place precisely fails
after executing $s$ but before the control has come back.
Indeed, consider $\{\dynat{p}{\{\spawned{s}\ \skip\}}\ \ t\}$ and
$\{\dynat{p}{\{\spawned{s}\}}\ \ t\}$. The local code $t$ is already
active only in the second statement while in the first one it is
waiting for the termination of the synchronous remote statement.
Accordingly, the second statement models the situation where the
control has come back locally after installing the remote asynchronous
computation.


As for error propagation, by rule {\sc (At)} we have that any
exception, either 
synchronous or asynchronous, that occurred remotely at place $p$ is
homomorphically reported locally at place $r$.
As an example, consider 
$\dynat{r}{\{\dynat{p}{\throw{\mathsf{E}}}\ \ \ t\}}$, 
then the exception at $p$ terminates the remote computation
and is reported at $r$ as a synchronous error so that to also discard
the local continuation $t$, whose execution depends on the completion
of the remote code. 
In order to recover from remote exceptions, we can use the try-catch
mechanism and write 
$\dynat{r}{\{\TryCatch{\,(\dynat{p}{\throw{\mathsf{E}}})\,}{t'}\ \ \  t\}}$
so that the synchronous exception is caught at $r$ according to the
rule {\sc (Try)}. 
More precisely, the $\try{s}{t}$ statement immediately activates $s$. 
Moreover, the rule {\sc (Try)} shows that asynchronous exceptions are
passed through, since they are only caught by \code{finish}.
On the other hand, 
synchronous exceptions are absorbed into a correct transition and the 
\code{catch}-clause is activated, together with the
(asynchronous) statements $s'$ left behind by the failed $s$. 

\subsection{Mechanization in Coq}
\label{subsec:x10-mechanization-coq}

We have encoded the syntax and semantics of TX10 in Coq, an
interactive theorem prover. 
Encoding the syntax and semantics are mostly straightforward, and
closely follows the paper presentation.  
However, the mechanized
formalism has a richer notion of exception propagation, which was
omitted from the paper for compactness.  Labels can carry a list of
exceptions, allowing multiple exceptions to be propagated by Finish
(instead of using a single generic exception).  Additionally, labels /
exceptions can be any value type. This complicates the rules, since
the (AT) rule needs to copy any values stored in the labels from the
target heap to the caller's heap. This is done by the actual X10
language, and correctly modeled by our mechanized semantics.

The most challenging part of encoding the semantics is encoding
the copy operation given in Definition~\ref{def:copy}, which copies an
object graph from one heap to another.

\subsubsection{Mechanizing the Copy Operation}
\label{subsec:copy}

Definition~\ref{def:copy} provides a declarative specification of the
copy operation, asserting the existence of a satisfying function.  The
mechanization explicitly constructs this function. In particular, it
provides a pure (provably terminating and side-effect free) function
with the given specification.

We first encode definitions of (local) reachability and graph
isomorphism, proving key theorems relating them.  We also define what
it means for a value to be \emph{well-formed} in a given heap: all objects
(locally) reachable from that value must be in the heap.  In other
words, the object graph rooted at the value may not contain dangling
pointers.

The implementation of the copy function itself proceeds recursively.
The recursive core of copy is given a list of existing mappings
(initially empty) from the source heap to the target heap, the source
and target heaps, and the initial object to copy.  For each field in
the object, if the value is an object identifier, it looks up the
identifier in the heap.  If heap does not contain the identifier
(which means that the given root is not well-formed), the copy
operation fails.  Otherwise, it creates a new object in the
destination heap, and adds a mapping from the the source oid to the
new oid.  It then calls itself recursively (with the enriched set
of mappings) to copy the object into the destination heap.  Finally,
the destination heap is updated so the newly created oid contains the
copied object returned by the recursive call.  The enriched set of
mappings is then returned so that it can be reused for the next field
in the object.

The tricky part of implementing this algorithm in Coq is proving
termination.  This is not obvious, since there can be cycles in the
object graph that we are copying. To prevent looping on such cycles,
the implementation carefully maintains and uses the set of existing
mappings from the source to the destination heap.
To prove termination for a non-structurally recursive function, we
define a well founded measure that provably decreases on every
recursive call. 
 We define this measure over the pair of the number of
oids in the source heap that are \emph{not} in the domain of the
mappings and the number of fields left in the object.  Since the
source heap is finite and does not change, this is a well founded
relation as long as as either the number of remaining elements goes
down (meaning that the number of distinct mappings increases) or it stays the
same and the number of fields decreases.

There are two recursive calls in the implementation.  The first
recursive call is during the processing of a field.  If the field
contains an oid, then the implementation adds a new pair to the set of
mappings before it calls itself.  The second recursive calls is part
of the iteration over the fields.  After processing a single field, it
calls itself recursively with the rest of them (without removing any
of the accumulated mappings).  In both cases, one of the measured
metrics decreases, ensuring that the recursive calls terminate.

As well as proving that the implementation is total, we also prove that
is has the required specification.  Moreover, if copy fails, there must
exist some oid reachable from the root which is not contained in the
heap.  This last part of the specification in turn enables us to prove
that copy will always succeed if the initial value is well formed.

\subsection{Properties of the transition relation}
\label{subsec:properties}

TX10 satisfies a number of useful properties, given below.  We have
mechanized these proofs in Coq, using our encoding of TX10.  This
provides a high level of assurance in these proofs, and fills in the
details of the various well-formedness conditions needed to ensure
that the properties hold.

\begin{proposition}[Absence of stuck states]\label{prop:no-stuck}
If a configuration $k$ is terminal then $k$ is of
the form $g$.
\end{proposition}

The mechanized proof of this proposition additionally proves that the
evaluation relation is computable: if the configuration is not
terminal, we can always compute a next step.  This is of course not
the only step, since the relation is non-deterministic.  Similarly, we
prove that the transitive closure of the evaluation relation does not
get stuck and is computable.  This proof can be ``run'',
yielding a simple interpeter for TX10.

\begin{definition}[Place-local Configuration]
Given a place-local heap $g$, we say that a configuration
$\conf{s}{g}$ is place-local if 
\begin{itemize}
\item for any local object id $o$ occurring in $s$ under
  \code{at}$(p)$ or $\dynat{p}{}$, we have that $o\in\dom{g(p)}$
  (hence $o\in\Loc_p$ by place-locality of $g$), and
\item for any global reference $o\$ q$ occurring in $s$, we have that
  $o\in\dom{g(q)}$.
\end{itemize}
\end{definition}

\begin{proposition}[Place-locality]
If $\conf{s}{g}$ is a place-local configuration and
$\conf{s}{g}\lambdared_p\conf{s'}{g'}~|~g'$, then 
$\conf{s'}{g'}$ is a place-local configuration, resp. $g'$ is a
place-local heap.
\end{proposition}

The following propositions deal with error propagation, whose
rationale can be summarized as follows:
{\em synchronous failures} arise from synchronous statements, and lead
to the failure of any synchronous continuation, while leaving
(possibly remote) asynchronous activities that are running in parallel
free to correctly terminate
(cf. Proposition~\ref{prop:only-failed-places=fail}).  
On the other hand, {\em asynchronous failures} arise when
an exception is raised in a parallel thread. In this case the
exception is confined within that thread, and 
it is caught by the closest $\finish{}$ construct that is waiting for
the termination of this thread. On termination of all spawned
activities, since one (or more) asynchronous exception were caught,
the $\finish{}$ constructs re-throws a synchronous failure
(cf. Proposition~\ref{prop:async-errors}).   
%
%
We rely on the following definition of {\em Evaluation Contexts}, that
is contexts under which a reduction step is possible:
$$
\begin{array}{lcl}
E & ::= & [\,] 
          ~\alt~ \{E\ t\} 
          ~\alt~ \{t\ E\} \mbox{ with } \vdash\isasync{t}
          ~\alt~ \dynat{p}{E}
\\[2mm]
 & &          ~\alt~ \spawned{E} ~\alt~ \MuFinish{E} 
           ~\alt~\TryCatch{E}{t}
\end{array}
$$

\begin{proposition}[Synchronous Failures]\label{prop:only-failed-places=fail}
If $\tuple{s,g} \VFAILderives_p k$
then
$\vdash \isSync{s}$. Moreover,
if $k\equiv\tuple{s',g'}$, then $\vdash \isAsync{s'}$
\end{proposition}

\begin{proposition}[Asynchronous Failures]\label{prop:async-errors}
\begin{itemize}
\item 
If $\conf{s}{g}\aFailred_p k$ then there exists an evaluation context
$E[\ ]$ such that $s=E[s_1]$ with $\conf{s'}{g}\aFailred_p k'$ and
$\vdash \isAsync{s_1}$.
%
\item 
If $\conf{\finishmu{\mu}{s}}{g}
\lred{\lambda_1}_p\ldots\lred{\lambda_n}_p g$ 
because of 
$\conf{s}{g} \lred{\lambda'_1}_p\ldots\lred{\lambda'_n}_p g$,
then
\begin{enumerate}
\item $\lambda_i=\epsilon$ for $i=1,\ldots,n-1$, and
\item 
  and $\lambda_n=\epsilon$ if $\forall j=1,\ldots, n \ $
  $\lambda'_j=\epsilon$  otherwise $\lambda_n=\mathsf{E}\otimes$.  
\end{enumerate}
\end{itemize}
\end{proposition}

\noindent
The proofs of the propositions above easily follow by induction on the
derivation of $\tuple{s,g} \VFAILderives_p k$, resp.
$\conf{s}{g}\aFailred_p k$, and an inspection of the rules for 
\code{finish}. 

\begin{proposition}\label{prop:AsyncRedux}
Let be $\conf{s}{g}\lambdared_p\conf{s'}{g'}$, then
if $\isAsync{s}$ then $\isAsync{s'}$, or equivalently,
if $\isSync{s'}$ then $\isSync{s}$
\end{proposition}


\section{Equivalence and Equational Laws}\label{sec:equational-laws} 

In this section we define a notion of equivalence for TX10 programs
along the lines of~\cite{x10-concur05}.
We consider weak bisimulation defined on both normal transitions and
transitions that throw an exception. Moreover, the bisimulation
encodes the observation power of the concurrent context in two ways:
($i$) it preserves the $\mathsf{isSync}/\mathsf{isAsync}$ predicate
and ($ii$) takes into account concurrent modification of shared
memory. As a result, we have that the resulting equivalence is a
congruence (cf. Theorem~\ref{thm:cong}).

We use a notion of {\em environment move} to model update of shared
heap by a concurrent activity. The store can be updated by updating a
field of an existing object, by creating a new (local) object, or by
means of a serialization triggered by a place shift. 

\begin{definition}[Environment move] 
An environment move $\Phi$ is a map on global heaps satisfying:
\begin{enumerate}
\item if $g$ is place-local, then $\Phi(g)$ is place-local, 
\item $\dom{\Phi(g)}=\dom{g}$, and $\forall p\in\dom{g}\ 
  \dom{g(p)}\subseteq\dom{\Phi(g)(p)}$.
\end{enumerate}
\end{definition}


Let $(\reduce_p)^*$ denote the reflexive and transitive
closure of $\lred{\epsilon}_p$, that is any number (possibly zero) of
$\epsilon$-steps. Then we let $\lRed{\lambda}_p$ stand for
$(\reduce_p)^*\lred{\lambda}_p(\reduce_p)^*$ when 
$\lambda\neq\epsilon$, and $(\reduce_p)^*$ if
$\lambda=\epsilon$.

\begin{definition}[Weak Bisimulation]\label{def:WeakBisim}
A binary relation ${\cal R}$ on closed configurations is a
{\em weak bisimulation} if whenever 
\begin{enumerate}
\item $g \ {\cal R}\ k$ then $k=g$,
\item $\tuple{s,g} \ {\cal R}\ k$ then $k=\tuple{t,g}$ for some $t$, 
  and
  \begin{itemize}
  \item $\vdash\issync{s}$ if and only if $\vdash\issync{t}$ and
  \item for every environment move $\Phi$, and
    for every place $p$ it is the case that  
  \begin{enumerate}
  \item if $\tuple{s,\Phi(g)} \Lambdaderives_p \tuple{s',g'}$ then for
    some $t'$, $\tuple{t, \Phi(g)}\lRed{\lambda}_p
    \tuple{t',g'}$ and $\tuple{s',g'}\ {\cal R}\ \tuple{t',g'}$,
    and vice versa.
  \item if $\tuple{s,\Phi(g)} \Lambdaderives_p g'$ then 
    $\tuple{t, \Phi(g)}\lRed{\lambda}_p g'$ and vice versa.
  \end{enumerate}
  \end{itemize}
\end{enumerate}
Two configurations are weak bisimilar, written
$\conf{s}{g}\equiv\conf{t}{g'}$, whenever there exists a weak
bisimulation relating them. The weak bisimilarity is the largest weak
bisimulation between configurations.
\end{definition}


\begin{theorem}\label{thm:cong}
Weak bisimilarity is a congruence.
\end{theorem}


We illustrate the equivalence by means of a number of equational laws
dealing with the main constructs of TX10.
{\small 
\begin{align}
\vdash \isSync{s} \quad\quad \{\Skip;\ s\} &\equiv s\\
\vdash \isSync{s} \quad\quad \{s\ \Skip;\} &\equiv s\\
\{\throw{v}\ \ s\} &\equiv \throw{v}\\
\{\{s\,t\}\ u\} &\equiv \{s\ \{t\,u\}\}
\end{align}
}
\noindent
To prove (1) it is sufficient to show that the relation
$
\{ (\conf{\{\skip\ s\}}{g}, \conf{s}{g}) ~|~ \vdash\isSync{s}\} \cup
Id_k 
$
where $Id_k$ is the identity relation over configurations, is a weak
bisimulation. 
Observe that (1) and (2)
only hold for synchronous statements since 
both $\{\skip\ s\}$ and $\{s\ \skip\}$ are
synchronous statements irrespective of $s$, hence the equivalence only
holds when also the r.h.s. is synchronous.

{\small \vspace*{-3mm}
\begin{align}
\TryCatch{\Skip}{t} &\equiv\Skip\\
\vdash \isSync{s} \quad\quad \TryCatch{\throw{v}}{s} &\equiv s\\
\TryCatch{s}{\throw{v}} &\equiv s\\
\TryCatch{\{s\,t\}}{u} & \not\equiv \{\TryCatch{s}{u}\ \ \TryCatch{t}{u}\}\\
\vdash \isAsync{\{s\,t\}} \quad\quad 
   \TryCatch{\{s\,t\}}{u} & \equiv \{\TryCatch{s}{u}\ \ \TryCatch{t}{u}\}\\
\TryCatch{(\TryCatch{s}{t})}{u} &\equiv \TryCatch{s}{(\TryCatch{t}{u})}
\end{align}
}
\noindent
Notice that law (8) is not valid, since the execution of the
r.h.s. might activate two copies of $u$ when both $s$ and $t$ fail in
sequence. On the other hand in the l.h.s. a synchronous error in $s$
implies that the continuation $t$ is discarded. Formally, when
$\conf{s}{g}\lred{v\otimes}_p g'$ then 
$\conf{\TryCatch{\{s\,t\}}{u}}{g} \reduce_p \conf{u}{g'}$ while the
r.h.s. reduces to $\conf{\{u \ \ \TryCatch{t}{u}\}}{g'}$.

{\small \vspace*{-3mm}
\begin{align}
\Atttwo{p}{\Skip} &\equiv \Skip \\
\Atttwo{p}{\throw{v}} &\equiv \throw{v} \\
  \Atttwo{p}{\{s\,t\}} & \equiv \ \{\Atttwo{p}{s}\ \ \Atttwo{p}{t}\}\\
\Atttwo{p}{(\TryCatch{s}{t})} &\equiv\TryCatch{(\Atttwo{p}{s})}{(\Atttwo{p}{t})}\\
\Atttwo{p}{\Atttwo{q}{s}} &\equiv \Atttwo{q}{s}
\end{align}
}
\noindent
All the laws above for place shift also hold for the dynamic version
of \code{at}. 

{\small \vspace*{-3mm}
\begin{align}
\spawned{\Skip} \not\equiv\Skip \quad\quad
\spawned{\throw{v}} &\not\equiv  \throw{v}\\
\{\spawned{\throw{v}}\ \ \spawned{\throw{v}}\} &
                                  \not\equiv \spawned{\throw{v}}
\end{align}
\begin{align}
\vdash\isAsync{s}\quad\quad 
  \{\spawned{\throw{v}}\ \ s\} &\equiv \{s\ \ \spawned{\throw{v}}\}\\
\async{\Atttwo{p}{s}} &\equiv \Atttwo{p}{\async{s}}\\
\async{\async{s}} &\equiv \async{s}\\
(\vdash \isAsync{s},\isAsync{t})\quad\quad\{s\,t\} &\equiv\{t\,s\}\\
(\vdash \isAsync{s})\quad\quad
   \TryCatch{\{s\,t\}}{u} &\equiv \{s\ \ \TryCatch{t}{u}\}
\end{align}
}
\noindent
Laws (16) do not hold since only the l.h.s. are asynchronous.
Law (17) does not hold since weak bisimilarity counts the number of
(asynchronous) exceptions, and the l.h.s. throws two asynchronous
$\mathsf{E}\times$ while the r.h.s. just one. Notice that by law (3)
we have $\{\throw{v}\ \throw{v}\}\equiv\throw{v}$, which is correct since
the l.h.s. throws a single $\mathsf{E}\otimes$ since
synchronous errors discard the continuation.

Observe that the static version of law (18) does not hold, i.e.,
$\{\async{\throw{v}}\ s\}\neq\{s\ \async{\throw{v}}\}$ since only in
the r.h.s. the statement $s$ can make a move. On the other hand, the
dynamic version of laws (19) and (20) are valid.
Law (21) comes observing that the relation
$
\{(\conf{\{s\ t\}}{g}, \conf{\{t\ s\}}{g})~|~
    \vdash \isAsync{s},\isAsync{t}\}\cup Id_k
$
is a weak bisimulation. Finally observe that law (22) only holds for
asynchronus $s$ since a synchronous error thrown by $s$ would be
caught in the l.h.s. while in the r.h.s. it would discard
the continuation.

{\small \vspace*{-3mm}
\begin{align}
\finish{\Skip} &\equiv\Skip\\
\finish{\throw{v}} &\not\equiv \throw{v}\\
\finish{\{s\ t\}} &\equiv \finish{s\ \ \finish{t}}\\
\finish{\{s\ \throw{v}\}} &\not\equiv \{\finish{s}\ \throw{v}\}\\
\finish{\async{s}} &\equiv \finish{s}\\
\finish{\{s\ \async{t}\}} &\equiv\finish{\{s\ t\}}
\end{align}
}
\noindent
Law (24) does not hold because of the exception masking
mechanism. More precisely, the exception $v\otimes$ thrown by
$\throw{v}$ is masked in the l.h.s. by $\mathsf{E}\otimes$ by the
finish construct.
For the same reason also law (26)
does not hold. Law (28) comes form (25) and (27).
In the following final set of laws we write $\vdash \noAsync{s}$ if
$s$ has no sub-term of the form $\async{s'}$ for some $s'$, i.e., if
$s$ cannot evolve to an asynchronous statement. 

{\small \vspace*{-3mm}
\begin{align}
\finish{\Atttwo{p}{s}} &\equiv\Atttwo{p}{\finish{s}}\\
\finish{\{\async{\throw{v}}\ \ s\}} &\equiv\{\finish{s}\ \ \throw{v}\}\\
\finish{\finish{s}} &\equiv \finish{s}\\
(\vdash \noAsync{s})\ \finish{s} &\not\equiv s\\
(\vdash \noAsync{s})\ \finish{\{s\,t\}} &\not\equiv \{s\,\finish{t}\}\\
(\vdash \noAsync{s})\ \finish{\TryCatch{s}{t}} &\not\equiv
\TryCatch{s}{\finish{t}}
\end{align}
}
\noindent
Again law (32), and then also (33) and (34), does not hold because of
the exception masking mechanism performed by the finish
construct.

\section{Resilient TX10}
\label{sec:resilientTX10}



The resilient calculus has the same syntax of TX10. We now assume that
any place $p\in\Places{\setminus}\{0\}$ can fail in any moment during
the program computation. Place $0$ has a special role: programs start
at place zero, then this place is used to communicate the result to
the user, so we assume it can never fail (if it does fail, the whole
execution is torn down).
%
In order to define the semantics, we now let global heaps $g$ to be
partial (rather than total) maps from places to local
heaps. Intuitively, $\dom{g}$ is the set of non failed places.
The semantics of Resilient TX10 is given by the rules in
Table~\ref{tab:ExprEval} and Table~\ref{tab:semBasic} from
Section~\ref{sec:tx10} plus the rules in
Tables~\ref{tab:semResI},~\ref{tab:semResII} and~\ref{tab:semResIII} 
given in this section. 
More precisely, the resilient calculus inherits form TX10 the rules
for expression evaluation (i.e., Table~\ref{tab:ExprEval})
and those in Table~\ref{tab:semBasic} which correspond to basic
statement executed at non-failed place $p$, i.e. $p\in\dom{g}$. 
The rules for TX10's main constructs, i.e. those in
Table~\ref{tab:semII}, hold also in the resilient calculus when 
$p\in\dom{g}$, but they must be integrated with additional rules
dealing with the case where the local place $p$ has failed.
Therefore, in order to improve the presentation, rather than
inheriting Table~\ref{tab:semII}, we collect here
all the operational rules for the main constructs, compacting them
in Tables~\ref{tab:semResI},~\ref{tab:semResII}
and~\ref{tab:semResIII}.

\begin{table}[t]
{\small
\begin{center}
\begin{tabular}{c}
\labeltree{(Place Failure)}
{p\in dom(g) }
{\conf{s}{g}\reduce_p\conf{s}{g\setminus\{(p,g(p))\}}}
\quad\quad
\labeltree{(Spawn)}
{}
{\begin{array}{ll}
 p\in\dom{g} & \conf{\async{s}}{g} \reduce_p \conf{\spawned{s}}{g} 
\\[1mm]
p\notin\dom{g} & \conf{\async{s}}{g} \sFailred{\DPE}_p  g
\end{array}
}
\\ \\
\labeltree{(Local Failure)}
{p\notin dom(g)}
{\begin{array}{rcl}
 \conf{\skip}{g} & \sFailred{\DPE}_p & g\\
  \conf{\throw{v}}{g} & \sFailred{\DPE}_p & g\\
 \conf{\ValS{x}{e}{s}}{g} & \sFailred{\DPE}_p & g \\
 \conf{e_1.f=e_2}{g}& \sFailred{\DPE}_p & g
 \end{array}
}
\quad\quad
\labeltree{(Async)}
{\conf{s}{g}\lambdared_p\conf{s'}{g'}~|~g'}
{\begin{array}{ll}
\lambda=\epsilon & 
     \conf{\spawned{s}}{g} \reduce_p \conf{\spawned{s'}}{g'}~|~g' 
\\[1mm]
\lambda=v\times,v\otimes\ \ &
    \conf{\spawned{s}}{g} \llred{{\mathsf{Msk}(v\times)}}{p} \conf{\spawned{s'}}{g'}~|~g'
\end{array}
}
\\ \\
\labeltree{(Finish)}
{\conf{s}{g}\lambdared_p\conf{s'}{g'}}
{\conf{\finishmu{\mu}{\!s}}{g} \reduce_p  \conf{\finishmu{\mu\cup\lambda}{\!s'}}{g'}}
\quad
\labeltree{(End of Finish)}
{\conf{s}{g}\lambdared_p g'
 \quad
\lambda'{=}\left\{\begin{array}{ll}
              \epsilon & \mbox{ if } \lambda{\cup}\mu{=}\emptyset\\
              {\mathsf E}\otimes & \mbox{ if }
            \lambda{\cup}\mu{\neq}\emptyset,
            p{\in}\dom{g}\\
             \DPE\otimes & \mbox{ if }
            \lambda{\cup}\mu{\neq}\emptyset, p{\notin}\dom{g} \\
              \end{array}
              \right.
}
{\conf{\finishmu{\mu}{s}}{g}\llred{{\mathsf{Msk}(\lambda')}}{p} g'}
%
\end{tabular}
\end{center}
}
\caption{Resilient Semantics I}
\label{tab:semResI}
\end{table}

\begin{table}[th]
{\small
\begin{center}
\begin{tabular}{c}
\labeltree{(Seq)}
{ \conf{s}{g}\lambdared_p\conf{s'}{g'}} 
{\begin{array}{ll}
\lambda=\epsilon,v\times & \conf{\{s\ t\}}{g} \lambdared_p
        \conf{\{s'\ t\}}{g'} 
\\[1mm]
\lambda=v\otimes & \conf{\{s\ t\}}{g} \lambdared_p
        \conf{s'}{g'} 
 \end{array}
}
\quad\quad
\labeltree{(Par)}
{\vdash\isasync{t}\quad \conf{s}{g}\lambdared_p\conf{s'}{g'}~|~ g'}
{\conf{\{t\ s\}}{g} \lambdared_p
        \conf{\{t\ s'\}}{g'}~|~\conf{t}{g'}}
\\ \\
\labeltree{(Seq Term)}
{p\in\dom{g}\quad \conf{s}{g}\lambdared_p g'}
{\begin{array}{ll}
\lambda=\epsilon,v\times & \conf{\{s\ t\}}{g} \lambdared_p
         \conf{t}{g'}
\\[1mm]
\lambda=v\otimes & \conf{\{s\ t\}}{g} \lambdared_p  g'
 \end{array}
}
\quad\quad
\labeltree{(Seq Failed Term)}
{p\notin\dom{g}\quad \conf{s}{g}\lambdared_p g'}
{\begin{array}{ll}
\vdash\isSync{s} & \conf{\{s\ t\}}{g} \lred{\DPE\otimes}_p g'       
\\[1mm]
\vdash \isAsync{s} & \conf{\{s\ t\}}{g} \lred{\DPE\times}_p  \conf{t}{g'}
 \end{array}
}
\end{tabular}
\end{center}
}
\caption{Resilient Semantics II}
\label{tab:semResII}
\end{table}

\begin{table}[Ht]
{\small
\begin{center}
\begin{tabular}{c}
\labeltree{(Place Shift)}
{(v',g')= \mathsf{copy}(v,q,g)}
{\begin{array}{ll}
p,q\in\dom{g} & \conf{\Att{q}{x\!}{\!v}{s}}{g} \reduce_p
  \conf{\dynat{q}{\{s[^{v'}\!/_x]\ \, \skip\}}}{g'}\\
q\notin\dom{g} & \conf{\Att{q}{x}{e}{s}}{g}  \sFailred{\DPE}_p  g
\\
p\notin\dom{g} & \conf{\Att{q}{x}{e}{s}}{g}  \sFailred{\DPE}_p  g
\end{array}
}
\\ \\
\labeltree{(At)}
{\conf{s}{g}\lambdared_q\conf{s'}{g'}~|~g'
 \quad
\lambda'{=}\left\{\begin{array}{ll}
             \DPE\otimes & \mbox{ if }
            \lambda{=}v\otimes,  p{\notin}\dom{g}\\
            \lambda & \mbox{ otherwise } 
              \end{array}
              \right.
}
{ \conf{\dynat{q}{s}}{g} \llred{{\mathsf{Msk}(\lambda')}}{p}  \conf{\dynat{q}{s'}}{g'}~|~g'}
\\ \\
\labeltree{(Try)}
{
 \conf{s}{g}\lambdared_p\conf{s'}{g'}~|~ g'}
{\begin{array}{ll}
\lambda=\epsilon,v\times & 
    \conf{\try{s}{t}}{g} \lambdared_p  \conf{\try{s'}{t}}{g'}~|~g'
\\[1mm]
p\in\dom{g},
\lambda=v\otimes & 
    \conf{\try{s}{t}}{g} \reduce_p  \conf{\{s'\  t\}}{g'}~|~\conf{t}{g'} 
\\[1mm]
p\not\in\dom{g},\lambda=\mathsf{D}\otimes & 
    \conf{\try{s}{t}}{g} \lambdared_p  \conf{s'}{g'}~|~g'
\end{array}     
}
\end{tabular}
\end{center}
}
\caption{Resilient Semantics III}
\label{tab:semResIII}
\end{table}

The place failure may occur at anytime, and it is modelled 
by the rule {\sc (Place Failure)} which removes the failed place from
the global heap. The semantics of TX10 is then extended so to
ensure that after the failure of a place $p$:
\begin{enumerate}
\item any attempt to execute a statement at $p$ results
  in a $\DPE$ exception (Proposition~\ref{prop:LocalFailDPE});
\item place shifts cannot be initiated form $p$
  nor launched to the failed $p$ (rule {\sc (Place Shift)});
\item any remote code that has been launched from $p$ before its
  failure is not affected and it is free to correctly terminate its
  remote computation. If a synchronous exception escapes from this
  remote 
  code and flows back at the failed place, then this exception is
  masked by a $\DPE$ (Proposition~\ref{prop:RemoteFailDPE}) which is
  thrown back to a parent \code{finish} construct waiting at a non
  failed place.
\end{enumerate}
More precisely, we will show that the operational semantics of
Resilient TX10 enforces the following three design principles: 
\begin{enumerate}
\item {\bf \em Happens Before Invariance Principle:}
failure of a place $q$ should not alter
the happens before relationship between statement 
instances at places other than $q$.
\item {\bf \em Exception Masking Principle:}
failure of a place $q$ will cause asynchronous exceptions thrown by 
  $\At{q}{s}$ statements to be masked by
  $\DPE$ exceptions.
\item {\bf \em Failed Place Principle:}
at a failed place, activating any statement or evaluating any
  expression should result in a $\DPE$ exception. 
\end{enumerate}

We now precisely illustrate the rules for the main constructs.
The rule {\sc (Local Failure)} shows that no expression is evaluated
at a failed place and any attempt to execute a basic
statement at the failed place results in a synchronous $\DPE$
exception. Similarly, rule {\sc (Spawn)} shows that new activities can
only be spawned at non failed places. On the other hand, rule 
{\sc (Async)} is independent form the failure of $p$, so that any
remote computation contained in $s$ proceeds not affected by the local
failure. The semantics of \code{finish} is the same as in
Section~\ref{sec:tx10} but for the rule {\sc (End of Finish)},
which now ensures that when $p\notin\dom{g}$ a $\DPE\otimes$ (rather
than $\mathsf{E}\otimes$) exception is thrown whenever one of the
governing activities (either local or remote) threw an exception.

The rules for sequences are collected in Table~\ref{tab:semResII}.
Rules {\sc (Seq)} and {\sc (Par)} are the same as in the basic
calculus, allowing remote computation under sequential or parallel
composition to evolve irrespective of local place failure.
The failure of $p$ plays a role only in rule {\sc (Seq Failed Term)}:
in this case the termination of the first component $s$ in the
sequence $\{s\ t\}$ always results in a $\DPE$ exception. Moreover, 
the continuation $t$ is discarded when $s$ is a synchronous statement.
On the other hand, when $s$ is an asynchronous statement, $t$ might be
an already active remote statement, hence the rule gives to $t$ the
chance to correctly terminate.


Rule {\sc (Place Shift)} allows the activation of a place-shift only
when both the source and the target of the migration are non-failed
places. Rule {\sc (At)} behaves like in TX10 except
that it masks any remote synchronous exception with a
\code{DeadPlaceException}. 
%
As an example consider
  $\dynat{p}{\{\dynat{q}{s}\ \dynat{r}{t}\}}$; if $p$ fails while
  $s$ and $t$ are (remotely) executing, it is important not to
  terminate the program upon completion of just $s$ (or just $t$).
  Then with rule {\sc (At)} we have that a remote computation silently
  ends even if the control comes back at a failed home.
As another example, consider
$\dynat{r}{\{\dynat{p}{\skip}\ \ t\}}$ with $p\notin dom(g)$, 
  then the failure of $\skip$ at $p$ must be reported at $r$
  as a synchronous error so that the continuation
  $t$ is discarded.

\begin{example}
Consider the following program, where the code $s_q$ is expected to be
executed at $q$ after the termination of any remote
activities recursively spawned at $p$: 
$$
\dynat{q}{\{\finish{\spawned{\dynat{p}{\{\finish{s}\ \ s_p\}}}}\ \ \ \ s_q\}}
$$ 
\noindent
Let also assume that $s$ spawns new remote activities running in a
third place $r$. Now, assume that both $p$ and $r$ fail while $s$ is
(remotely) executing. We have that $s$ throws an exception that should
be detected by the inner \code{finish}, however since $p$ is a failed
place, termination and error detection in $s$ must be delegated to the
outer \code{finish} waiting at non failed place $q$: 
that is indeed performed by rule {\sc (End of Finish)}.
Hence we have that the \code{finish} at $q$ throws a synchronous error
and the continuation $s_q$ is discarded. 
Notice that enclosing the inner $\finish{}$ within a try-catch
construct is only useful when $p$ is a non failed place. Indeed,
consider the program
$$
\dynat{q}{\{\finish{\spawned{\dynat{p}{\{\TryCatch{(\finish{s})}{t}\ \ s_p\}}}}\ \ \ \ s_q\}}
$$ 
then by the rule {\sc (Try)} for exception handling
we have that when $p$ is a failed place the clause is never
executed, hence he two programs above have the same semantics.
On the other hand, we can recover from an exception in $s$ by
installing a try/catch at the non failed place $q$:
$
\dynat{q}{\{ \TryCatch{(
    \finish{\spawned{\dynat{p}{\{\finish{s}\ \ s_p\}}}})}{t}\ \ \ \ s_q\}}
$. 
\end{example}


\subsection{Properties of the transition relation}


The main properties of the operational semantics of TX10 scale to
Resilient TX10.  We have encoded the syntax and semantics of Resilient
X10 in Coq, as we did for TX10 (see
Section~\ref{subsec:x10-mechanization-coq}).  Using this encoding, we
have mechanized the analogous proofs for Resilient X10.

\begin{proposition}[Absence of stuck states]\label{prop:no-stuckRES}
If a configuration $k$ is terminal then $k$ is of
the form $g$.
\end{proposition}

The definition of place-locality of configurations must be generalized
to the case of partially defined heaps. More precisely, given a
configuration $\conf{s}{g}$, any local oid $o$ is $s$ must be locally
defined, while a global reference $o\$ p$ might now be a dangling
reference since the global object's home place $p$ might have failed.

\begin{definition}[Place-local Resilient Configuration]
Given a place-local heap $g$, we say that a configuration
$\conf{s}{g}$ is place-local if $\forall p\in\dom{g}$
\begin{itemize}
\item for any local object id $o$ occurring in $s$ under
  \code{at}$(p)$ or $\dynat{p}{}$, we have that $o\in\dom{g(p)}$
  (hence $o\in\Loc_p$ by place-locality of $g$).
\end{itemize}
\end{definition}

Given the definition above, we can still prove that resilient
semantics preserves place-locality of resilient configurations.
\begin{proposition}[Place-locality]
If $\conf{s}{g}$ is a place-local resilient configuration and
$\conf{s}{g}\lambdared_p\conf{s'}{g'}~|~g'$, then 
$\conf{s'}{g'}$ is a place-local resilient configuration, resp. $g'$
is a place-local heap.
\end{proposition}

Also Proposition~\ref{prop:only-failed-places=fail} and
\ref{prop:async-errors} hold also in Resilient TX10, with a minor
modification: in the second clause of
Proposition~\ref{prop:async-errors} the final error thrown by a
\code{finish} construct might be either $\mathsf{E}\otimes$ or
$\DPE\otimes$. 

The main results of this section are the three principles stated
above. The Exception Masking Principle, formalized by
Theorem~\ref{thm:EMP}, shows that no exception other than $\DPE$ can
arise form a failed place. The Failed Place Principle, formalized by
Theorem~\ref{thm:FPP}, shows that no statement can be executed at a
failed place. Finally, the Happens Before Invariance Principle shows
in Theorem~\ref{thm:HBIP} that the place failures do not alter the
happens before relation between the non-failed statements.

\begin{theorem}[Exception Masking Principle]\label{thm:EMP}
Let be $p\notin\dom{g}$ and $\conf{s}{g}\lambdared_p k$. 
If $\lambda=v\otimes$, then $v=\DPE$. 
\end{theorem}

Let say $\vdash \isLocal{s}$ whenever $s$ does not contain active
remote computation, that is $s$ has no substatements of the form
$\dynat{q}{s'}$. 
We say $\vdash\isRemote{p}{s}$ when any basic statement in $s$ occurs
under a $\dynat{q}{}$ construct for some place $q$ with $q\neq p$.

\begin{proposition}[Local failure]\label{prop:LocalFailDPE}
Let be $p\notin\dom{g}$ and $\conf{s}{g}\lambdared_p k$.
\begin{itemize}
\item
If $\vdash \noAsync{s}$ and $\vdash\isLocal{s}$, then 
$\lambda=\DPE\otimes$ and $k=g$.
\item If $\vdash\isLocal{s}$, then either
\begin{itemize}
\item $\lambda=\DPE\otimes$ or $\lambda=\DPE\times$, or 
\item $s=E[\finishmu{\mu}{t}]$, $\conf{s}{g}\reduce_p\conf{s'}{g'}
        \Reduce_p\lred{\DPE\otimes}_p g'$
      with $s'=E[\finishmu{\DPE}{t'}]$
 and $\conf{t}{g}\llred{\DPE\otimes\ or\ \DPE\times}{p}\conf{t'}{g'}$.
\end{itemize}
\end{itemize}
\end{proposition}

The following proposition states that remote computation at non-failed
place proceeds irrespective of local place failure, but for the
exception masking effect.

\begin{proposition}[Remote computation]\label{prop:RemoteFailDPE}
Let be $\vdash\isRemote{p}{s}$. If $\conf{s}{g}\lambdared_p
\conf{s'}{g'}~|~g'$ with $p\in\dom{g}$, then  
$\conf{s}{g}\reduce_p\conf{s}{g\setminus\{(p,g(p))\}}
  \llred{\lambda'}{p} \conf{s'}{g'_*}~|~g'_*$ where
$g'_*=g'\setminus\{(p,g'(p))\}$ and
$\lambda'=\lambda$ if $\lambda=\epsilon,v\times$ while
$\lambda'=\DPE\otimes$,  if 
$\lambda=v\otimes$. Moreover
$\vdash\isRemote{p}{s'}$. 
\end{proposition}

\begin{theorem}[Failed Place Principle]\label{thm:FPP}
If $s$ performs a correct step at a failed place, i.e.,
$\conf{s}{g}\reduce_p\conf{s'}{g'}~|~g'$ with $p\notin\dom{g}$, 
then either
\begin{itemize}
\item $s$ contains a substatement that remotely computed a correct
  step at a non failed place, i.e., $s=E[s_1]$ with
  $\vdash\isRemote{p}{s_1}$,
  $\conf{s_1}{g}\reduce_p\conf{s'_1}{g'}~|~g'$ and $s'=E[s'_1]$,
  or
\item a local activity ends at $p$ with a $\DPE$ that has been
  absorbed by a governing finish, i.e. 
  $s=E[\MuFinish{t}]$, $s'=E[\finishmu{\DPE}{t'}]$ and
  $\conf{t}{g}\llred{\DPE\otimes\ or\ \DPE\times}{p}\conf{t'}{g'}$.
\end{itemize}
\end{theorem}

\noindent
We denote by $\vec{k}$ a trace 
$\conf{s_0}{g_0}\lred{\lambda_1}_0\conf{s_1}{g_1}
\lred{\lambda_2}_0\ldots \lred{\lambda_n}_0\conf{s_n}{g_n}$. 
Moreover we write $|\vec{k}|$ for the length $n$ of such
a trace, and $k_i$ to indicate the $i$-th configuration
$\conf{s_i}{g_i}$, $i=0,...,n$. We define below the {\em Happens
  Before} relation in terms of the operational semantics. 
Intuitively, given a program $s$ with two substatements
$s_1,s_2$, we say that $s_1$ happens before $s_2$ whenever in any
program execution $s_1$ is activated, i.e. it appears under an
evaluation context, before $s_2$.
We refer to~\cite{yuki2013array} for a static definition of the
happens before relation in terms of static statements, which is also
proved to be equivalent to a dynamic characterization that
correspond to the following one.

\begin{definition}[Happens Before]
Let 
$s_0$ be a program and let $s_1, s_2$ be two
substatements of $s_0$. Then we say that $s_1$ {\em happens before}
$s_2$, written $s_1<s_2$, whenever
for any trace $\vec{k}$ such that $k_0=\conf{s_0}{g_0}$ and
$k_{|\vec{k}|}=\conf{E[s_2\rho]}{g}$ for some $g$, some evaluation
context $E$ and some variable substitution $\rho$, there exists $i\in
0,...,|\vec{k}|$ such that $k_i=\conf{E'[s_1\rho']}{g'}$ for some
$g',E',\rho'$. 
\end{definition}

Notice that the definition of the Happens Before relation is
parametric on a transition relation. Let write $s_1<s_2$ when
we restrict to (traces in) TX10 semantics, and $s_1<_{R}s_2$ when
considering (traces in) the resilient semantics. 
%
%

\begin{theorem}[Happens Before Invariance]\label{thm:HBIP}
Let 
$s_0$ be a program and let $s_1, s_2$ be two
substatements of $s_0$. 
Then $s_1<s_2$ if and only if $s_1<_{R}s_2$.
\end{theorem}

\subsection{Equational laws}

The equational theory of TX10 can be smoothly generalized to the
resilient calculus.
In order to scale the notion of weak bisimilarity to
Resilient TX10 we have to consider generalized environment moves that
take into account the failure of a number of places. 

\begin{definition}[Resilient Environment move] 
An environment move $\Phi$ is a map on global heaps satisfying:
\begin{enumerate}
\item if $g$ is place-local, then $\Phi(g)$ is place-local, 
\item $\dom{\Phi(g)}\subseteq\dom{g}$, and 
$\forall p\in\dom{\Phi(g)}\ \dom{g(p)}\subseteq\dom{\Phi(g)(p)}$.
\end{enumerate}
\end{definition}

The weak bisimilarity for Resilient TX10 is then defined as in 
Definition~\ref{def:WeakBisim}, where we rely on resilient environment
moves and the operational steps used in the bisimulation game are
those defined in this section. In particular, this means that also
place failures occurring at any time must be simulated by
equivalent configurations. 
We discuss in the following which laws are still valid in the
Resilient calculus. 

{\small \vspace*{-3mm}
\begin{align}
\vdash \noAsync{s} \quad\vdash \isLocal{s}\quad \{\Skip;\ s\} &\equiv s\\
\vdash \isSync{s} \quad\quad\{s\ \Skip;\} &\not\equiv s\\
\{\throw{v}\ \ s\} &\equiv \throw{v}\\
\{\{s\,t\}\ u\} &\equiv \{s\ \{t\,u\}\}\\
\vdash\noAsync{s}\ \vdash\isLocal{s}\quad
                 \TryCatch{\throw{v}}{s} &\equiv s
\end{align}
}
In order for law (1) of Section~\ref{sec:tx10} to be valid also at a
failed place, law (35) requires a stronger constraint for $s$ so to 
ensure that in that case also the r.h.s always throw a $\DPE\otimes$.
On the other hand law (36) never holds since the failure of the local
place can happen after the completion of $s$ but before the execution
of $\Skip$, thus only the l.h.s. would throw a $\DPE\otimes$.
As for the Try/Catch constructs, all the rules of TX10 are still
valid, but for rule (6), which must be substituted by law (39). 
Indeed, similarly to law (35), we must ensure that $s$
throws a synchronous $\DPE\otimes$ error whenever the local place is
failed. 

{\small \vspace*{-3mm}
\begin{align}
\Atttwo{p}{\Skip} &\not\equiv \Skip \quad\quad\quad
\dynat{p}{\Skip}\ \not\equiv\ \Skip\\
\Atttwo{p}{\throw{v}} &\equiv \throw{v} \quad\quad
  \dynat{p}{\throw{v}} \ \equiv\ \throw{v}\\
  \Atttwo{p}{\{s\,t\}} & \not\equiv \ \{\Atttwo{p}{s}\ \ \Atttwo{p}{t}\}\\
\dynat{p}{\{s\,t\}} &\equiv \{\dynat{p}{s}\ \ \dynat{p}{t}\}\\
\Atttwo{p}{(\TryCatch{s}{t})} &\not\equiv\TryCatch{(\Atttwo{p}{s})}{(\Atttwo{p}{t})}\\
\Atttwo{p}{\Atttwo{q}{s}} &\not\equiv \Atttwo{q}{s}\quad\quad
  \dynat{p}{\dynat{q}{s}}\ \not\equiv \ \dynat{q}{s}
\end{align}
}
\noindent
The laws (40) for place shift does not hold in the resilient
calculus since they involve two terms that run in different places
that might fail in different moments. Notice that law (41) is still
valid by means of the exception masking principle. Rule (42) does not
hold anymore since the local place can fail after the completion of
$s$ but before the place shift of $t$. On the other hand law (43) is
still valid since both terms already run at the same place $p$ and the
failure of local place does not affect remote computation.
Law (44) does not hold anymore since $p$ may fail after $s$ has thrown
an exception but before the activation of the handling $t$. 
The first law (45) does not hold since $p$ may fail before the
place-shift at $q$, while the second law does not hold since the
failure of $p$ would mask any exception thrown at $q$.

{\small \vspace*{-3mm}
\begin{align}
\vdash\isAsync{s}\quad\quad 
  \{\spawned{\throw{v}}\ \ s\} &\equiv \{s\ \ \spawned{\throw{v}}\}\\
\async{\Atttwo{p}{s}} &\not\equiv \Atttwo{p}{\async{s}}\\
\spawned{\dynat{p}{s}} &\not\equiv \dynat{p}{\spawned{s}}\\
\async{\async{s}} &\equiv \async{s}\\
(\vdash \isAsync{s},\isAsync{t})\quad\quad\{s\,t\} &\equiv\{t\,s\}\\
(\vdash \isAsync{s})\quad\quad
   \TryCatch{\{s\,t\}}{u} &\equiv \{s\ \ \TryCatch{t}{u}\}\\
\finish{\Atttwo{p}{s}} &\not\equiv\Atttwo{p}{\finish{s}}
\end{align}
}
Laws (47) (48) does not hold anymore because of the exception masking
effect. Indeed, if $s$ remotely throws a synchronous exception
$v\otimes$, we have that the r.h.s. throws a $v\times$ exception while
the l.h.s. throws $\DPE\otimes$ by means of masking. 

%

All the laws for finish hold also in Resilient TX10 but for the one
involving place shifting. In law (52) a difference appears between the
two terms when the remote place $p$ fails after the remote code has
been activated. In this case $s$ throws a $\DPE$ exception at the
failed place, but in the l.h.s. the local (non failed) \code{finish}
masks this exception as a generic $\mathsf{E}$, while in the
r.h.s. the exception reported locally is still $\DPE$.

\section{Conclusions and Future work}
\label{sec:conclusion}

We have studied a formal small-step structural operational semantics
for TX10, that is a large fragment of the X10 language covering
multiple places, shared mutable objects, sequences, \code{async},
\code{finish}, \code{at} and \code{try/catch} constructs. 
We have then shown that this framework smoothly extends to the
case where places dynamically fail. Failure is exposed through
exceptions thrown by any attempt to execute a statement at the failed
place. The error propagation mechanism in Resilient TX10 extends that
of TX10 ($i$) by discarding exception handling at failed places,
i.e. no \code{catch} clause is ever executed at failed places, and
($ii$) by masking with a {\em DeadPlaceException} any remote exception
flowing back at the failed place. Moreover, we established a Happens
Before Invariance Principle showing that the failure of a place $p$
does not alter the happens before relationship between statements at
places other than $p$.

As an example of formal methods that can be developed on top of the
given operational semantics, we studied a bisimulation based
observation equivalence. We showed that it correctly encodes the
observation power of the concurrent context by proving that it is a
congruence. We illustrated this equivalence by means of a number of
laws dealing with the main constructs of the language, discussing
which of these equivalences are invariant under place failures.
The axiomatization of the given equivalence is left for future work.
We think that the resilient equational theory opens the way to the
development of laws that can be used in the X10 compiler to optimize
programs, e.g. using polyhedral analysis \cite{yuki2013array}.
We also plan for future work the extension of the framework we
presented to cover the \code{atomic} and \code{when} constructs from
X10.  
We also plan to develop denotational semantics for TX10 based on a
pomset model that naturally allows
the definition of the happens before relation.
Another promising approach seems to be the study of full abstraction
by extending to this setting the trace set model of
S.~Brookes \cite{Brookes93fullabstraction}.



\begin{thebibliography}{10}

\bibitem{rmi-oopsla05}
Alexander Ahern and Nobuko Yoshida.
\newblock Formalising java rmi with explicit code mobility.
\newblock In {\em OOPSLA '05}, pages 403--422, New York, NY, USA, 2005. ACM.

\bibitem{millwheel}
Tyler Akidau, Alex Balikov, Kaya Bekiroglu, Slava Chernyak, Josh Haberman,
  Reuven Lax, Sam McVeety, Daniel Mills, Paul Nordstrom, and Sam Whittle.
\newblock {MillWheel: Fault-Tolerant Stream Processing at Internet Scale}.
\newblock In {\em Very Large Data Bases}, pages 734--746, 2013.

\bibitem{amadio-failure}
Roberto~M. Amadio.
\newblock An asynchronous model of locality, failure, and process mobility.
\newblock In {\em Coordination Languages and Models}, LNCS, pages 374--391.
  Springer, 1997.

\bibitem{coqart}
Y.~Bertot and P.~Cast{\'e}ran.
\newblock {\em Interactive Theorem Proving and Program Development: Coq'Art:
  The Calculus of Inductive Constructions}.
\newblock Texts in Theoretical Comp. Sci. Springer, 2004.

\bibitem{MJ}
G.M. Bierman, M.J. Parkinson, and A.~M. Pitts.
\newblock Mj: An imperative core calculus for java and java with effects.
\newblock Technical report, University of Cambridge Computer Laboratory, 2003.

\bibitem{failureOfFailures}
Frank S.~de Boer, Joost~N. Kok, Catuscia Palamidessi, and Jan J. M.~M. Rutten.
\newblock The failure of failures in a paradigm for asynchronous communication.
\newblock In {\em CONCUR '91}, pages 111--126, London, UK, UK, 1991.
  Springer-Verlag.

\bibitem{Brookes93fullabstraction}
Stephen Brookes.
\newblock Full abstraction for a shared variable parallel language.
\newblock In {\em In Proceedings, 8th Annual IEEE Symposium on Logic in
  Computer Science}, pages 98--109. IEEE Computer Society Press, 1993.

\bibitem{brookes-CSL}
Stephen Brookes.
\newblock A semantics for concurrent separation logic.
\newblock {\em Theor. Comput. Sci.}, 375(1-3):227--270, April 2007.

\bibitem{x10-oopsla05}
Philippe Charles, Christian Grothoff, Vijay Saraswat, Christopher Donawa, Allan
  Kielstra, Kemal Ebcioglu, Christoph von Praun, and Vivek Sarkar.
\newblock X10: an object-oriented approach to non-uniform cluster computing.
\newblock In {\em OOPSLA '05}, pages 519--538, New York, NY, USA, 2005. ACM.

\bibitem{upc2005upc}
UPC Consortium et~al.
\newblock {UPC} language specifications.
\newblock {\em Lawrence Berkeley National Lab Tech Report LBNL--59208}, 2005.

\bibitem{ResilientX10}
David Cunningham, David Grove, Benjamin Herta, Arun Iyengar, Vijay Saraswat,
  Olivier Tardieu, Kiyokuni Kawachiya, Hiroki Murata, and Mikio Takeuchi.
\newblock {Resilien X10: Efficient failure-aware programming}.
\newblock In {\em POPL}, 2014.

\bibitem{scale-graph}
Miyuru Dayarathna, Charuwat Houngkaew, and Toyotaro Suzumura.
\newblock Introducing {S}calegraph: an {X10} library for billion scale graph
  analytics.
\newblock In {\em X10 '12}, pages 6:1--6:9, New York, NY, USA, 2012. ACM.

\bibitem{dean-mr}
Jeffrey Dean and Sanjay Ghemawat.
\newblock Mapreduce: Simplified data processing on large clusters.
\newblock In {\em OSDI'04}, pages 10--10, Berkeley, CA, USA, 2004. USENIX
  Association.

\bibitem{Fournet96acalculus}
C{\'e}dric Fournet, Georges Gonthier, Jean-Jacques L{\'e}vy, Luc Maranget, and
  Didier R{\'e}my.
\newblock A calculus of mobile agents.
\newblock In {\em CONCUR '96}, pages 406--421, London, UK, UK, 1996.
  Springer-Verlag.

\bibitem{DBLP:journals/iandc/FrancalanzaH08}
Adrian Francalanza and Matthew Hennessy.
\newblock A theory of system behaviour in the presence of node and link
  failure.
\newblock {\em Inf. Comput.}, 206(6):711--759, 2008.

\bibitem{DPI}
Matthew Hennessy.
\newblock {\em A Distributed Pi-Calculus}.
\newblock Cambridge University Press, New York, NY, USA, 2007.

\bibitem{lee2010feather}
Jonathan~K. Lee and Jens Palsberg.
\newblock Featherweight {X10}: a core calculus for async-finish parallelism.
\newblock In {\em PPoPP '10}, pages 25--36, New York, NY, USA, 2010. ACM.

\bibitem{pregel}
Grzegorz Malewicz, Matthew~H. Austern, Aart~J.C Bik, James~C. Dehnert, Ilan
  Horn, Naty Leiser, and Grzegorz Czajkowski.
\newblock Pregel: A system for large-scale graph processing.
\newblock In {\em Proceedings of the 2010 ACM SIGMOD International Conference
  on Management of Data}, SIGMOD '10, pages 135--146, New York, NY, USA, 2010.
  ACM.

\bibitem{Riely:2001:DPL:504563.504586}
James Riely and Matthew Hennessy.
\newblock Distributed processes and location failures.
\newblock {\em Theor. Comput. Sci.}, 266(1-2):693--735, September 2001.

\bibitem{X10}
Vijay Saraswat, Bard Bloom, Igor Peshansky, Olivier Tardieu, and David Grove.
\newblock {X10} language specification version 2.2, March 2012.
\newblock \url{x10.sourceforge.net/documentation/languagespec/x10-latest.pdf}.

\bibitem{x10-concur05}
Vijay Saraswat and Radha Jagadeesan.
\newblock Concurrent clustered programming.
\newblock In {\em CONCUR 2005 - Concurrency Theory}, pages 353--367, London,
  UK, 2005. Springer-Verlag.

\bibitem{Shinnar:2012:MIP:2367502.2367513}
Avraham Shinnar, David Cunningham, Vijay Saraswat, and Benjamin Herta.
\newblock {M3R: increased performance for in-memory Hadoop jobs}.
\newblock {\em Proc. VLDB Endow.}, 5(12):1736--1747, August 2012.

\bibitem{gml}
{X10} {G}lobal {M}atrix {L}ibrary.
\newblock \url{https://x10.svn.sourceforge.net/svnroot/x10/trunk/x10.gml},
  October 2011.

\bibitem{yuki2013array}
Tomofumi Yuki, Paul Feautrier, Sanjay Rajopadhye, and Vijay Saraswat.
\newblock Array dataflow analysis for polyhedral x10 programs.
\newblock In {\em POPL'13}, 2013.

\bibitem{Zaharia:2010:SCC:1863103.1863113}
Matei Zaharia, Mosharaf Chowdhury, Michael~J. Franklin, Scott Shenker, and Ion
  Stoica.
\newblock Spark: cluster computing with working sets.
\newblock In {\em HotCloud'10}, pages 10--10, 2010.

\end{thebibliography}


\end{document}